\documentclass[11pt, a4paper]{article}

\usepackage{jheppub}

\usepackage{amssymb,amsmath,amsfonts}
\usepackage{subfigure}
\usepackage{mathtools}
\usepackage{braket}

\newcommand{\beq}{\begin{equation}}
\newcommand{\eeq}{\end{equation}}
\newcommand{\bea}{\begin{eqnarray}}
\newcommand{\eea}{\end{eqnarray}}
\newcommand{\amp}{{\cal{M}}}

\newcommand{\dd}{\text{d}}
\newcommand{\bs}{\boldsymbol}
\newcommand{\F}{{\cal F}}
\newcommand{\Hint}{{\cal H}}
\newcommand{\ev}[1]{\langle #1 \rangle}

\begin{document}

\preprint{\begin{flushright}MAN/HEP/2013/25\\
    IPPP/13/101\\
    DCPT/13/202
  \end{flushright}}
\arxivnumber{1312.3871}

\title{Manifest causality in quantum field theory \\
  with sources and detectors}

\author[a]{Robert Dickinson,}
\author[a]{Jeff Forshaw,}
\author[a,b]{Peter Millington,}
\author[c]{and Brian Cox}
\affiliation[a]{
Consortium for Fundamental Physics, School of Physics \& Astronomy,\\
University of Manchester, Manchester M13 9PL. U.K.}
\affiliation[b]{
Institute for Particle Physics Phenomenology, Durham University,\\
Durham DH1 3LE. U.K.}
\affiliation[c]{
School of Physics \& Astronomy,
University of Manchester,\\  Manchester M13 9PL. U.K.}

\emailAdd{robert.dickinson@hep.manchester.ac.uk}
\emailAdd{jeff.forshaw@manchester.ac.uk}
\emailAdd{peter.millington@hep.manchester.ac.uk}
\emailAdd{brian.cox@hep.manchester.ac.uk}

\abstract{  We introduce  a way  to compute  scattering  amplitudes in
quantum field theory including  the effects of particle production and
detection. Our amplitudes are manifestly causal, by which we mean that
the  source and detector  are always  linked by  a connected  chain of
retarded propagators. We show how these amplitudes can be derived from
a    path   integral,    using    the   Schwinger-Keldysh    ``in-in''
formalism. Focussing on $\phi^3$  theory, we confirm that our approach
agrees with  the standard  S-matrix approach in  the case  of positive
energy plane-wave scattering.  }
  
\keywords{Quantum Field Theory, Thermal Field Theory}

\maketitle
\flushbottom

\section{Introduction}

Although relativistic quantum field  theories are built with causality
in mind,  the way  causality plays  out at the  level of  the particle
dynamics  is  not so  clear.  The  usual  ``in-out'' formalism  places
production and  detection sources on  the same footing  and amplitudes
involve only  the Feynman propagator. Our approach  replaces the usual
expression for the scattering matrix by an ``in-in'' expectation value
in which detection is distinct from production. It leads to manifestly
causal results, in which the retarded progagator is prominent.

In  section \ref{sec:expectation}  we  begin by  presenting a  general
result for the expectation value  of a hermitian operator that is some
local function of  field operators; we have in  mind that the operator
represents an observable.   We do this in the  presence of an external
source and  show that the effect  of the source is  transmitted to the
detector  via  an  unbroken  chain of  retarded  propagators.  Feynman
propagators necessarily  appear but  never so as  to break  the causal
link from source to detector.

In  section  \ref{sec:scatter}  we   present  an  expression  for  the
calculation of scattering amplitudes.  Since these can be expressed as
expectation values  of non-local products  of field operators,  we can
use many of the results of  the previous section.  We then present the
corresponding  Feynman rules  and illustrate  their use  in tree-level
scattering.

In  section   \ref{sec:pathintegral}  we  show   that  our  scattering
amplitudes  can be  derived  from  a path  integral  in the  ``in-in''
formalism. In  section \ref{sec:smatrix}, we prove  the equivalence of
our scattering  amplitudes, in the case of  positive energy plane-wave
scattering,  with  those  obtained  from  the  S-matrix.  In  appendix
\ref{sec:cutting},  we describe  the link  with the  cutting  rules of
thermal field theory.

\section{Expectation values of local operators}
\label{sec:expectation}

We     shall    consider     a    single     real     scalar    field,
$\phi_x\:\equiv\:\phi(x)$,   in   the   presence  of   some   external
disturbance that is active  within a spacetime region $\mathcal{R}_J$.
We suppose that  a measurement taking place within  a spacetime region
$\mathcal{R}_B$  is represented by  a hermitian  operator $B$  that is
some function of the field  operator within the region, i.e.~$B \equiv
B(\{\phi_x:  x\in\mathcal{R}_B\})$,   and  that  the   disturbance  is
represented by a contribution to the Hamiltonian of the form
$$ \int_{{\mathcal{R}}_J} \dd^4 x \; \gamma {\cal{J}}(\phi_x,x)~.$$

Given  a  source  whose  strength  is parametrised  by  $\gamma$,  the
sensitivity  $\sigma_B$ of a  detector making  measurement $B$  may be
expressed as
\begin{equation}
  \sigma_B\!=\!\frac{\partial}{\partial\gamma}\ev{B}.
  \label{eq:sensitivity}
\end{equation}
In what follows,  we will show that $\sigma_B$  can be expressed using
chains    of   retarded    propagators    from   $\mathcal{R}_J$    to
$\mathcal{R}_B$, which implies that it vanishes outside of the forward
light cone of the source.

In the interaction picture (which we always employ) the system evolves
as
\begin{equation}
  \ket{\psi(t)} = U(t,t_A)\ket{\psi_A} 
\end{equation}
where 
\begin{align}
  U(t,t_A) = 1+(-i)\!\!\int_{t_A}^t \!\dd t_1\, H_1 &+
  (-i)^2\!\!\int_{t_A}^t \dd t_1 \dd t_2\,\Theta_{12}\, H_1 H_2
  + \ldots \nonumber\\
  & + (-i)^n\!\!\int_{t_A}^t \dd t_1\ldots \dd t_n\,
  \Theta_{1\ldots n}\,H_1 \dots H_n + \ldots~,
\end{align}
is    the     evolution    operator    and    $\Theta_{ijk...}\!\equiv
\Theta(t_i\!>\!t_j\!>\!t_k\!>\!\ldots)$   is  a   Heaviside  function,
defined to be  1 if the time-ordered condition  within the brackets is
satisfied  and  0 otherwise.  $H_i  \equiv  H_\text{int}(t_i)$ is  the
interaction Hamiltonian, including the effect of the source.

The expectation  value of an  operator $B$ at  a given time  $t_0$ may
then be written as
\begin{align}
  \ev{B}_{t=t_0} &= \bra{\psi_A}\; U^\dag(t_0,t_A)\,B(t_0)\,U(t_0,t_A)
  \;\ket{\psi_A}.
  \label{eq:evb}
\end{align}
Rather than express $U(t,t_A)$ as a time-ordered exponential, we use a
generalisation        of       the        Baker--Hausdorff       lemma
\cite{franson2002perturbation}, i.e.~with $B_0\equiv B(t_0)$ we write
\begin{equation}
  U^{\dag}(t_0,t_A)\,B_0\,U(t_0,t_A)=F_0+ F_1 + F_2 +\ldots
\end{equation}
with each term expressed as a set of nested commutators:
\begin{align}
  F_0 &= B_0 \nonumber\\
  F_1 &= (-i)\!\!\int_{t_A}^{t_0} \dd t_1 \,\Theta_{01}\big[B_0,H_1\big]
  \nonumber\\
  F_2 &= (-i)^2\!\!\int_{t_A}^{t_0} \dd t_1 \dd t_2 \,\Theta_{012}
  \Big[\big[B_0,H_1\big],H_2\Big] \nonumber\\
  &\vdots\nonumber\\
  F_n &= (-i)^n\!\!\int_{t_A}^{t_0}\dd t_1 ... \dd t_n \,\Theta_{01\ldots n}
  \bigg[...\bigg[\Big[\big[B_0,H_1\big],H_2\Big],H_3\bigg]...,H_n\bigg]
  \nonumber\\
  &\vdots
  \label{eq:nestedcoms}
\end{align}
Note           that           within          each           integral,
$t_0\!>\!t_1\!>\!\ldots\!>\!t_n\!>\!t_A$. To encompass the full extent
of the  influence of  the source, $t_A$  must be  a time prior  to any
point in $\mathcal{R}_J$.

For the expectation value of $B$, we have
\begin{align}
  \ev{B}_{t=t_0} &= \sum_{n=0}^\infty \bra{\psi_A}\,F_n\,\ket{\psi_A}
  \nonumber\\
  \text{where}\;\;\;\; F_n &= (-i)^n\!\!\int\dd^4x_1 ... \dd^4x_n \,\
  \Theta_{01\ldots n}\,\mathcal{F}_n,
  \nonumber\\
  \text{and}\;\;\;\;\;\;\; \mathcal{F}_n &=
  \big[\mathcal{F}_{n\!-\!1},\mathcal{H}_n\big] \;\;\;
  \text{with}\;\;\mathcal{F}_0 \equiv B_0~,
  \label{eq:xnspacetime}
\end{align}
where     $H_\text{int}(t)   =    \int    \dd^3   \boldsymbol{x}    \;
\mathcal{H}(x)$.  Each  term in the $n$th  order perturbation operator
$\mathcal{F}_n$   involves  $n$   spacetime   points  $x_i$,   located
progressively further back in time as $i$ increases.

An expression for the commutator  of functions of field operators with
the Pauli-Jordan  function defined by  $[\phi_j,\phi_i] = \Delta_{ij}$
can be found  in \cite{transtrum2005commutation}. For our perturbation
series,
\begin{align}
  \big[\mathcal{F}_{r\!-\!1}&(\phi_0,\phi_1,\ldots,\phi_{r\!-\!1})\,,\,
  \mathcal{H}(\phi_r) \big] \nonumber\\
  &=-\underbrace{\sum_{k_0=0}^\infty\sum_{k_1=0}^\infty \ldots
    \!\!\sum_{k_{r\!-\!1}=0}^\infty}_{K_r\equiv\sum_{i=0}^{r\!-\!1}\!k_i\;\neq\,0}
  \!\left(\prod_{i=0}^{r\!-\!1}\frac{(\Delta_{ir})^{k_i}}{k_i!}\right)
  D_0^{k_0}D_1^{k_1} \cdots D_{r-1}^{k_{r-1}}
  \mathcal{F}_{r\!-\!1}\,
  D_{r}^{K_r} \, \mathcal{H} \nonumber\\
  &\equiv \mathcal{F}_{r}(\phi_0,\phi_1,\ldots,\phi_{r})~.
  \label{eq:transtrum}
\end{align}
The  operator $D_i$  is  defined via  $D_i[ \phi_0^{n_0}  \phi_1^{n_1}
\cdots  \phi_i^{n_i}  \cdots  ]  = (\phi_0^{n_0}  \phi_1^{n_1}  \cdots
n_i\phi_i^{n_i-1}  \cdots)$  and  the  sums  $k_i$ run  from  zero  to
infinity,  with   the  exclusion  of   the  case  in  which   all  are
simultaneously    zero.    Every   one    of   the    $n$   iterations
$(r\!=\!1,\ldots,n)$   required  to   generate   $\mathcal{F}_n$  from
$\mathcal{F}_0$ generates  a set of  terms, each of which  contains at
least one  factor of  $\Delta_{ir}$ for some  $i\!<\!r$. Each  term in
$\mathcal{F}_n$  therefore  refers  to   a  set  of  spacetime  points
$x_0,x_1,\ldots,x_i,\ldots,x_r,\ldots,x_n$ in which every member $x_i$
is  connected  to at  least  one earlier  point  $x_r$  by a  retarded
propagator.   Thus  every  non-zero  contribution to  the  sensitivity
$\sigma_B$ of the detector must  contain an unbroken chain of retarded
propagators from a point in the source.

\paragraph{Phi-cubed theory.}

As  an   illustrative  example  we   shall  suppose  that   $B(t_0)  =
\phi(t_0,\boldsymbol{x}_0)=\phi(x_0)$ and
$$ \Hint(x) = \frac{g}{3!}\phi_x^3 -\gamma J_x\phi_x~.$$
We assume  that the system  can be approximated  by the vacuum  of the
non-interacting   theory   at   $t    =   -\infty$,   i.e.~we   take
$\ket{\psi_A}=\ket{0}$.

The first-order commutator is
\begin{align}
  \mathcal{F}_1=\big[\phi_0,\mathcal{H}_1\big] &
  = \frac{g}{3!}\big[\phi_0,\phi_1^3\big]-\gamma J_1\big[\phi_0,\phi_1\big]
  \nonumber\\
  &= -\bigg(\frac{g}{2}\phi_1^2-\gamma J_1\bigg) \Delta_{01}~,
\end{align}
where $J_i\equiv J(x_i)$. This gives
\begin{align}
  F_1 &= -i\!\int\dd^4x_1 \,\bigg(\frac{g}{2}\phi_1^2-\gamma J_1\bigg)
  \,\Delta^{\text{R}}_{01}~,
  \label{eq:f1op}
\end{align}
where   $\Delta^{\text{R}}_{ij}\equiv   \Delta_{\text{R}}(x_i,x_j)   =
-\Theta(x^0_i\!-\!x^0_j)\Delta_{ij}$  is the  retarded  propagator for
the  free  field.\footnote{The  relevant  propagators are  defined  in
appendix  \ref{app:props}.} Note the  relative minus  sign due  to our
definition of $\Delta_{ij}=[\phi_j,\phi_i]$.

The                       second-order                      commutator
$\mathcal{F}_2=\big[\mathcal{F}_1,\mathcal{H}_2\big]$ is then
\begin{align}
  \label{eq:comm2}
  \mathcal{F}_2 &= -\frac{g^2}{12} \Delta_{01} \big[\phi_1^2,\phi_2^3\big]
  +\gamma J_2\frac{g}{2}\Delta_{01}\big[\phi_1^2,\phi_2\big]
  \nonumber\\
  &= \frac{g}{2}\bigg\{ g\big(\phi_1\phi_2^2+\phi_2 \Delta_{12}\big)
  -2\gamma J_2\phi_1 \bigg\} \Delta_{12}\Delta_{01}~.
\end{align}

Finally, the third-order commutator is
\begin{align}
  \mathcal{F}_3 &= \frac{g}{2}\bigg\{
  -g^2\bigg(\phi_1\phi_2\phi_3^2+\phi_1\phi_3\Delta_{23}
  +2\phi_2\phi_3\Delta_{13}+\Delta_{23}\Delta_{13}
  +\frac{1}{2}\phi_3^2\Delta_{12}\bigg)\Delta_{23}
  \nonumber\\
  &\hspace*{1cm}-\frac{1}{2}g^2\phi_2^2\phi_3^2\Delta_{13}
  + g\gamma J_2\phi_3^2\Delta_{13}
  +g\gamma J_3\bigg(2\phi_1\phi_2\Delta_{23}+\phi_2^2\Delta_{13}
  + \Delta_{23}\Delta_{12}\bigg)\nonumber\\
  &\hspace*{1cm}-2\gamma^2 J_2 J_3\Delta_{13}\bigg\}
  \Delta_{12}\Delta_{01}.
  \label{eq:comm3}
\end{align}
We may  now evaluate  the expectation value  of $B$.  Any uncontracted
field operators can  be handled using Wick's Theorem  and give rise to
Feynman propagators:
\begin{equation}
  \Delta^{\text{F}}_{ij}\equiv \Delta_{\text{F}}(x_i,x_j) =
  \bra{0}\text{T}[\phi_i\phi_j]\ket{0}~.
\end{equation}
It then follows that the first term in the expectation value expansion
eq.~(\ref{eq:xnspacetime}) is, from eq.~(\ref{eq:f1op}),
\beq
  \bra{0} F_1 \ket{0} = -i\!\int\dd^4x_1 \,\bigg(\frac{g}{2}
  \Delta^{\text{F}}_{11}-\gamma J_1\bigg) \Delta^{\text{R}}_{01}~.
  \label{eq:first}
\eeq

\begin{figure}[h]
  \centering
  \subfigure[]{\raisebox{0\textheight}
    {\includegraphics[height=0.12\textheight]{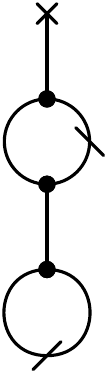}}}
  \enskip
  \subfigure[]{\raisebox{0\textheight}
    {\includegraphics[height=0.12\textheight]{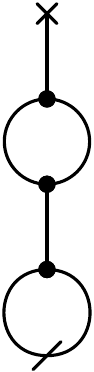}}}
  \enskip
  \subfigure[]{\raisebox{0.02\textheight}
    {\includegraphics[height=0.1\textheight]{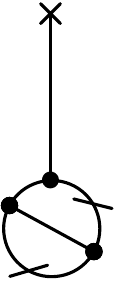}}}
  \enskip
  \subfigure[]{\raisebox{0.02\textheight}
    {\includegraphics[height=0.1\textheight]{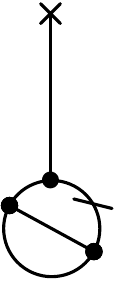}}}
  \enskip
  \subfigure[]{\raisebox{0.02\textheight}
    {\includegraphics[height=0.1\textheight]{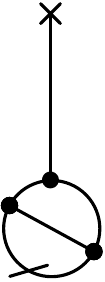}}}
  \enskip
  \subfigure[]{\raisebox{0.02\textheight}
    {\includegraphics[height=0.1\textheight]{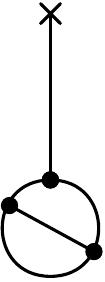}}}
  \enskip
  \subfigure[]{\raisebox{0.02\textheight}
    {\includegraphics[height=0.1\textheight]{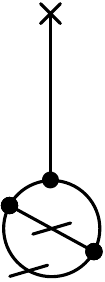}}}
  \enskip
  \subfigure[]{\raisebox{0.0\textheight}
    {\includegraphics[height=0.12\textheight]{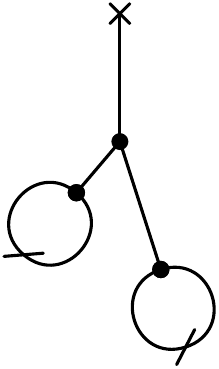}}
    \label{fig:f31}}
  \enskip
  \subfigure[]{\raisebox{0.02\textheight}
    {\includegraphics[height=0.1\textheight]{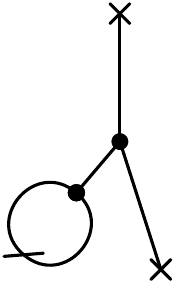}}
    \label{fig:f32}}
  \enskip
  \subfigure[]{\raisebox{0.02\textheight}
    {\includegraphics[height=0.1\textheight]{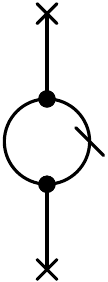}}}
  \enskip
  \subfigure[]{\raisebox{0.02\textheight}
    {\includegraphics[height=0.1\textheight]{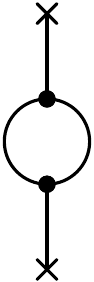}}}
  \enskip
  \subfigure[]{\raisebox{0.02\textheight}
    {\includegraphics[height=0.1\textheight]{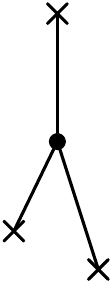}}
    \label{fig:f33}}
    \caption{The  diagrams corresponding  to  $\bra{0}F_3\ket{0}$. The
    co-ordinate  $x_0$ is labelled  by the  cross at  the top  of each
    graph  and the lower  points are  at $x_1$,  $x_2$ and  $x_3$. The
    time-ordering of $x_2$  and $x_3$ in diagrams (h),  (i) and (l) is
    no longer fixed by  retarded propagators. Retarded propagators are
    represented by unslashed lines  and Feynman propagators by slashed
    lines.}
  \label{fig:f3x}
\end{figure}

The second term, from  eq.~(\ref{eq:comm2}), involves an odd number of
fields    and    therefore   vanishes.    The    third   term,    from
eq.~(\ref{eq:comm3}) and depicted in figure \ref{fig:f3x}, is
\bea
  \bra{0} F_3 \ket{0} 
  &=& ig\!\int\dd^4x_1\dd^4x_2\dd^4x_3 \bigg\{
  g^2\bigg(\frac{1}{2}\Delta^{\text{F}}_{12}\Delta^{\text{F}}_{33}
  +\Delta^{\text{F}}_{13}\Delta^{\text{F}}_{23}
  -\frac{1}{2}\Delta^{\text{F}}_{13}\Delta^{\text{R}}_{23}
  -\Delta^{\text{F}}_{23}\Delta^{\text{R}}_{13}
  \nonumber\\
  & &+\frac{1}{2}\Delta^{\text{R}}_{23}\Delta^{\text{R}}_{13}
  -\frac{1}{4}\Delta^{\text{F}}_{33}\Delta^{\text{R}}_{12}\bigg)
  \Delta^{\text{R}}_{23}
  +g^2\bigg(\frac{1}{8}\Delta^{\text{F}}_{22}\Delta^{\text{F}}_{33}
  +\frac{1}{4}(\Delta^{\text{F}}_{23})^2 \bigg)
  \Delta^{\text{R}}_{13}
  \nonumber\\
  & &-g\gamma
  J_3\bigg(\frac{1}{2}\Delta^{\text{F}}_{22}\Delta^{\text{R}}_{13} +
  \Delta^{\text{F}}_{12}\Delta^{\text{R}}_{23} -
  \frac{1}{2}\Delta^{\text{R}}_{23}\Delta^{\text{R}}_{12}\bigg) +
  \frac{1}{2}\gamma^2 J_2
  J_3\Delta^{\text{R}}_{13}\bigg\}\Delta^{\text{R}}_{12}
  \Delta^{\text{R}}_{01}~.
  \label{eq:f3x}
\eea
Of            particular             note            is            the
$\Delta^{\text{F}}_{22}\Delta^{\text{F}}_{33}$ term on the second line
(which corresponds  to figure  \ref{fig:f31}). This includes  an extra
factor  of  $1/2$ because  there  is  a  residual Heaviside  function,
$\Theta_{23}$,   that   cannot   be   absorbed   into   the   retarded
propagators. However, because of  the symmetry under interchange of $2
\leftrightarrow  3$  we  can  drop  the  time-ordering  constraint  in
exchange for a symmetry factor of $1/2!$.  The same Heaviside function
is  also  present  in  the  term  $\propto J_2  \,  J_3$  (see  figure
\ref{fig:f33})  and  its  elimination  also  gives rise  to  a  factor
$1/2$.  For  these two  diagrams,  we  say that  points  2  and 3  are
`equivalent'.                       Finally,                       the
$\Delta^{\text{F}}_{22}\Delta^{\text{R}}_{13}\, J_3$ term on the final
line (see  figure \ref{fig:f32})  also originally had  a $\Theta_{23}$
which  we eliminated  by  combining it  with  a contribution  $\propto
\Delta^{\text{F}}_{33}\Delta^{\text{R}}_{12}\,  J_2$. In this  way all
explicit  time-ordering Heaviside functions  disappear from  the final
expression.

From the first few nested commutators, it is evident that the majority
of  the   terms  in  $\bra{0}B\ket{0}$  are   vacuum  diagrams,  which
contribute nothing to the sensitivity of the detector to the source.

It  is also  apparent  that there  is  a straightforward  relationship
between the set of diagrams that can  be drawn and the form of a given
nested commutator. The  rules are listed below for  the case $B(t_0) =
\phi(t_0)$.

\begin{figure}[h]
  \centering
  \subfigure[]{\raisebox{0\textheight}
  {\includegraphics[height=0.07\textheight]{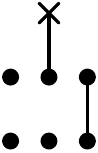}}}
  \qquad \qquad
  \subfigure[]{\raisebox{0\textheight}
  {\includegraphics[height=0.07\textheight]{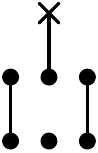}}}
  \qquad \qquad
  \subfigure[]{\raisebox{0\textheight}
  {\includegraphics[height=0.07\textheight]{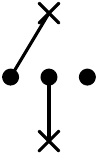}}}
  \caption{The three  contributions to  $\F_2$. A horizontal  row of
  dots represents a vertex  and any uncontracted dots are understood
  as field  operators.  The  graphs have an  associated combinatoric
  and coupling factor: for (a) and (b) this is $18 (g/3!)^2$ and for
  (c) it is $6(g/3!)(-\gamma J_2)$. }
  \label{fig:f2dot}
\end{figure}

\paragraph{Feynman rules.} To  compute $\F_n$ we are to  draw a set of
skeleton graphs  involving the  $n+1$ times from  $t_0$ to  $t_n$. The
graphs  can be  built up  iteratively starting  from the  latest time,
$t_0$, and ending with the earliest  time $t_n$. We are to draw either
a  cubic interaction vertex  (arising from  $g\phi^3/3!$) or  a source
term (arising from  $\gamma J_n \phi$) for all  times $t_1$ to $t_{n}$
inclusive. For each time earlier than $t_0$ there must be at least one
retarded  propagator heading  forwards in  time. This  means  that the
vertices may  include uncontracted field operators (i.e.~there can be
fewer than 3  retarded propagators at any vertex  but never zero). For
each vertex we  associate a factor of $g/3!$ and  a factor of $\phi_i$
for  each  uncontracted   field  operator  (written  in  chronological
order). For each instance of the  source there is a factor of $-\gamma
J_i$ and  there is  an overall factor  of $(-1)^n$. For  each retarded
propagator  between  $x_j$ (earlier)  and  $x_i$  (later) associate  a
factor of $\Delta_{ij}$. There is a combinatoric factor for the number
of  different ways  to contract  the  fields in  forming the  retarded
propagators.  This is the  prescription to compute the operator $\F_n$
and   it  is   illustrated  for   the   case  of   $\F_2$  in   figure
\ref{fig:f2dot}.

To compute  $\bra{0} F_n \ket{0}$  we take each skeleton,  convert the
$\Delta_{ij}$  to  $-\Delta^{\text{R}}_{ij}$  and compute  the  vacuum
Green's functions associated with the incomplete vertices using Wick's
Theorem. There is a  factor $-i\int \dd^4 x_i$ for all $i  > 0$ and we
must take  care to absorb  the time-ordering Heaviside  functions into
the retarded propagators: if  any Heaviside functions remain then they
can be eliminated provided we associate a symmetry factor of $1/m!$ if
there are  $m$ `equivalent' spacetime  points (in the  sense explained
above). For example, there is a factor $1/2$ for graphs (h) and (l) in
figure \ref{fig:f3x}.

\section{Scattering amplitudes}
\label{sec:scatter}

\begin{figure}[h]
  \begin{center}
    \includegraphics[width=0.32\textwidth,angle=0]{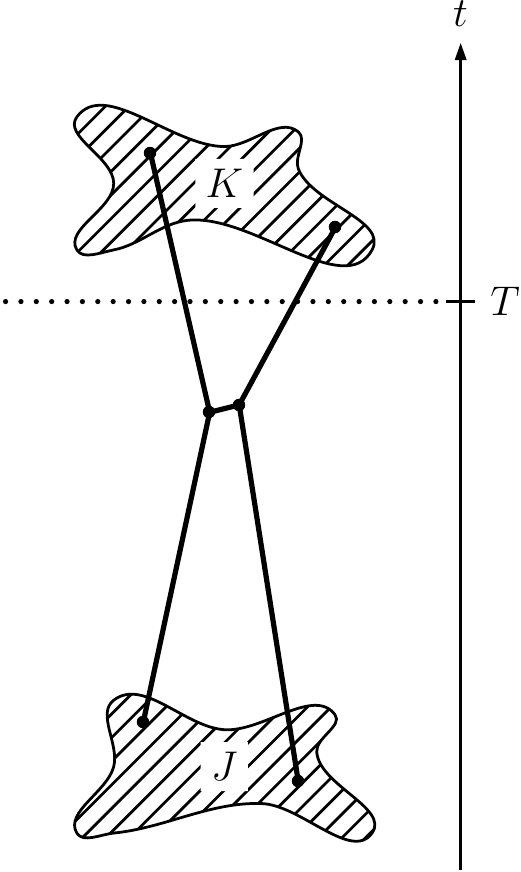}
    \caption{Source-to-detector scattering.\label{fig:scatter}}
  \end{center}
\end{figure}

We use  the following expression  for the scattering amplitude  in the
presence of a detector source $K_x$ and a production source $J_x$:
\begin{eqnarray}
  \Gamma_{JK}  & = & \langle T | \left[ \text{T} \exp \left(i
  \int_{T}^{\infty} \dd^4 x
  \; K_x\phi_x \right) \right] | T \rangle 
  \label{eq:master}
\end{eqnarray}
where $| T \rangle  \equiv U(T,-\infty) | 0 \rangle$ and
\beq
  U(t',t) \equiv \text{T} \exp \left( -i\int_{t}^{t'}
  \dd^4 x \,  (H_\text{int}(x) -  J_x \phi_x )\right)~.
\eeq
Again we take
\beq
  H_\text{int}(x) = \frac{g}{3!}\phi_x^3 ~.
\eeq
We will  assume that $K_x$ has support  only in the future  of $T$ and
that the  cubic interaction  and source $J_x$  are turned off  in that
region.  The  situation is  illustrated  in figure  \ref{fig:scatter}.
Equation  (\ref{eq:master}) looks just  like an  ``in-in'' expectation
value and as such it inherits  many of the properties described in the
previous section.  Eq.~\eqref{eq:master} is a  scattering amplitude in
the sense that $\exp(i\int_T^{\infty} \dd^4 x \,K_x \phi_x)$ is rather
like  an S-matrix  operator for  determining the  future of  the state
$|T\rangle$. For  simplicity, we have  assumed that the  detector acts
locally, e.g.~there are no bilocal terms $\sim K(x,y)\phi(x)\phi(y)$.

In the case of $n$-to-two scattering, we can extract the relevant part
of the amplitude, $\Gamma_{JK}^{n \to 2}$, from
\bea
  \sum_{n=0}^{\infty} \; \Gamma_{JK}^{n \to 2} &=&
  -\int \dd^4 x \, \dd^4y \; \frac{1}{2} K_xK_y\langle 0 |
  U^\dagger(T,-\infty) \text{T} [ \phi_x
  \phi_y]  U(T,-\infty) |0 \rangle\;.
  \label{eq:mab}
\eea
As in the  previous section, and since we are  assuming $x^0,y^0 > T$,
eq.~(\ref{eq:mab})  can  be  re-arranged by  commuting  $\phi_x\phi_y$
through the  time-evolution operator using  the Baker-Hausdorff lemma,
i.e.
\beq
  U^\dagger(T,-\infty) \phi_x \phi_y U(T,-\infty) = \sum_n F_n
\eeq
where
\beq
  F_n = (-i)^n \prod_{j=1}^n \int \dd^4 x_j \; \Theta_{T1\cdots n} \,
  \F_n
\eeq
and
\beq
  \F_n = [\F_{n-1},\Hint_n] ~~~~\text{with}~~~\F_0 =
  \phi_x \phi_y~~\text{and}~~\Hint_n = g\phi_n^3/3! - J_n \phi_n~.
\eeq
To  compute  two-to-two scattering  at  tree-level  we  would need  to
compute the part of $\F_4$ that is proportional to $J^2 g^2$.

The amplitude  can also  be obtained using  the same Feynman  rules as
articulated in the  previous section. The only difference  is that the
field operator that we are  averaging can be a non-local polynomial of
the field.  This merely  introduces extra points,  all later  than the
time $T$, at  which propagators may terminate. For  each such point we
will have a factor of $-i\int \dd^4 x_i ~ K_i$.

\paragraph{Some examples  at tree-level.} We begin  by considering the
one-to-two amplitude, which is
\begin{align}
  \label{eq:1to2}
  \Gamma^{1\to2}_{JK}\ &=- g\prod_{j=1}^2 \left(\int \dd^4 x_j \right)
  \int\!\mathrm{d}^4x\int\!\mathrm{d}^4y\;\Theta_{xT}\;\Theta_{yT}\;
  \Theta_{T1}\;
  \nonumber\\&\qquad \qquad \qquad\qquad
  \times\:\frac{1}{2}K_xK_yJ_2\bigg[
  \Delta^{\text{R}}_{x1}\bigg(\Delta^{\text{F}}_{y1}
  -\frac{1}{2}\Delta^{\text{R}}_{y1}\bigg)
  \Delta^{\text{R}}_{12} + ( x \leftrightarrow y )\bigg]~.
\end{align}
Notice  that this  includes  a Feynman  propagator  coupling from  the
interaction vertex to  the detector. There is however  no violation of
causality  because the measurement  is the  coherent detection  of two
particles at  points $x$ and $y$,  which is causal since  one of those
particles is constrained to lie  in the future lightcone of the source
by  the unbroken  chain of  retarded  propagators. In  fact we  should
anticipate  such   superfically  acausal  correlations:   they  are  a
manifestation of entanglement.

\begin{figure}[h]
  \begin{center}
    \includegraphics[height=0.18\textwidth]{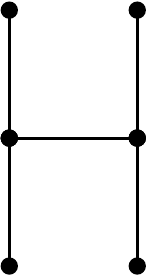}\qquad
    \includegraphics[height= 0.18\textwidth]{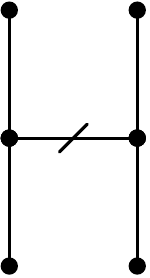}\qquad
    \includegraphics[height=0.18\textwidth]{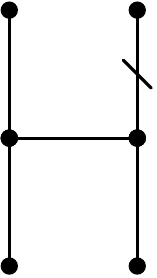}\qquad
    \includegraphics[height=0.18\textwidth]{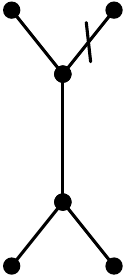}\qquad
    \includegraphics[height=0.18\textwidth]{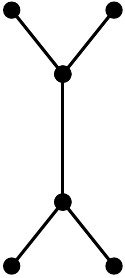}
  \end{center}
    \caption{The five graphs relevant for 2-to-2 scattering.  Retarded
    propagators  are  represented   by  unslashed  lines  and  Feynman
    propagators by slashed lines.\label{fig:2to2}}
\end{figure}

We will now consider two-to-two  scattering. In this case, the Feynman
rules give
\bea
  \Gamma^{2\to 2}_{JK} &=& - g^2 \prod_{j=1}^4 \left(\int \dd^4 x_j \right)
  \int \dd^4 x \, \dd^4y \;  \Theta_{xT} \Theta_{yT} \; \Theta_{T1} \;
  \frac{1}{2}K_xK_yJ_3J_4
  \nonumber \\ & & \qquad \qquad 
  + \bigg \{ [ \Delta^{\text{R}}_{x1}\bigg(\Delta^{\text{F}}_{y1}
  -\frac{1}{2}\Delta^{\text{R}}_{y1}\bigg) \Delta^{\text{R}}_{12}
  \Delta^{\text{R}}_{23} \Delta^{\text{R}}_{24} + ( x \leftrightarrow y )]
  \nonumber \\
  & & \qquad \qquad \qquad \qquad
  + \{ [ \Delta^{\text{R}}_{x1} \Delta^{\text{R}}_{13} \Delta^{\text{R}}_{y2}
  \Delta^{\text{R}}_{24} (
  \Delta^{\text{F}}_{12} \, \Theta_{12}
  - \Delta^{\text{R}}_{12}) + ( 1 \leftrightarrow 2)] 
  + [ 3 \leftrightarrow 4]  \}
  \nonumber \\ & &  \qquad \qquad \qquad \qquad
  + \{ [ \Delta^{\text{R}}_{x1} \Delta^{\text{R}}_{13} \Delta^{\text{F}}_{y2}
  \Delta^{\text{R}}_{24} + (1 \leftrightarrow2)]
  \Delta^{\text{R}}_{12}  + [ 3  \leftrightarrow 4] \}
  \bigg\}~.
  \label{eq:2to2}
\eea
The  first line  in the  large  braces of  eq.~(\ref{eq:2to2}) is  the
$s$-channel contribution,  and the first  set of terms in  brackets on
each  of  the  third   and  fourth  lines  generates  the  $t$-channel
contribution. The  $u$-channel contribution  is obtained from  the $[3
\leftrightarrow 4]$ interchange  of the $t$-channel contribution. Note
the residiual  $\Theta_{12}$ on  the second line.  It combines  with a
$\Theta_{21}$   term  after   interchange  $(1   \leftrightarrow  2)$.
Equation  (\ref{eq:2to2}) can be  simplified somewhat  by symmetrizing
the sources:
\bea \label{eq:2to2sym}
  \Gamma^{2\to2}_{JK} &=&  - g^2 \prod_{j=1}^4 \left(\int \dd^4 x_j \right)
  \int \dd^4 x \, \dd^4y \;  \Theta_{xT} \Theta_{yT} \; \Theta_{T1}
  K_x K_y J_3 J_4
  \\ &  &  \qquad \qquad 
  \times \bigg \{  
  \frac{1}{4}\Delta^{\text{R}}_{x1}(2\Delta^{\text{F}}_{y1}
  -\Delta^{\text{R}}_{y1}) \Delta^{\text{R}}_{12}
  \Delta^{\text{R}}_{23} \Delta^{\text{R}}_{24}
  \nonumber\\& & \qquad \qquad \qquad \qquad
  + \frac{1}{2}\Delta^{\text{R}}_{x1} \Delta^{\text{R}}_{13}
  \Delta^{\text{R}}_{y2} \Delta^{\text{R}}_{24} 
  (\Delta^{\text{F}}_{12} - 2\Delta^{\text{R}}_{12}) 
  + \Delta^{\text{R}}_{x1} \Delta^{\text{R}}_{13} \Delta^{\text{F}}_{y2}
  \Delta^{\text{R}}_{24} \Delta^{\text{R}}_{12}
  \bigg\}\;.
  \nonumber
\eea 
Written in this form, there are only 5 distinct graphs to consider and
these are illustrated in figure \ref{fig:2to2}.

We can make explicit contact with the corresponding S-matrix amplitude
for the  scattering of momentum  eigenstates by promoting  the sources
$K$ and $J$ to operators in  Fock space.  Specifically, if we take $K$
and $J$ to be replaced by
\begin{align}
  \label{eq:pwsources}
  K_x\ &\to \phi^{\mathrm{out}}_x (\Box_x^2\:+\:m^2)\;,
  \nonumber \\
  J_x\ & \to \phi^{\mathrm{in}}_x(\Box_x^2\:+\:m^2)\;,
\end{align}
and  we  take  the limit  $T\:\to\:\infty$  in  such  a way  that  the
integrals over $x^0$ and $y^0$ can be approximated by integration over
an   infinite   time  domain,   which   allows   the  preparation   of
freely-propagating  momentum eigenstates at  $t\:\to\:\pm\:\infty$. We
may then take the  overlap of eq.~(\ref{eq:2to2sym}) with two-particle
``in'' and ``out'' states, i.e.
\begin{equation}
  \amp_{JK}^{2 \to 2} \ \equiv\
  \braket{\mathrm{out};\,\mathbf{k}_3,\mathbf{k}_4|\Gamma_{JK}^{2 \to 2}|
  \mathrm{in};\,\mathbf{k}_1,\mathbf{k}_2}\;.
\end{equation}
After  expressing  the  5  propagators  in  eq.~(\ref{eq:2to2sym})  in
momentum  space,  the spacetime  integrals  can  be performed  leaving
behind an overall energy-momentum conserving delta function. The final
result is as expected:
\bea
  \label{eq:smat}
  \amp^{2\to2}_{JK} &=& -g^2 \; (2\pi)^4 \delta^{(4)}(k_1+k_2-k_3-k_4) \;
  \left( 
  \frac{i}{s - m^2} + 
  \frac{i}{t - m^2} +
  \frac{i}{u - m^2} \right)~, 
\eea
where $s = (k_1 + k_2)^2$, $t = (k_1-k_2)^2$ and $u = (k_1-k_3)^2$ are
the  Mandelstam variables.   To  illustrate the  point,  the last  two
graphs  of figure \ref{fig:2to2},  which correspond  to the  first two
terms in the second  line of eq.~(\ref{eq:2to2sym}), contribute to the
$s$-channel amplitude with weights $2-1 = 1$.

\begin{figure}[h]
  \centering
  {\raisebox{0\textheight}
  {\includegraphics[height=0.13\textheight]{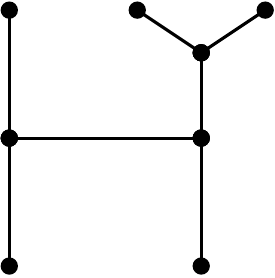}}}
  \qquad 
  {\raisebox{0\textheight}
  {\includegraphics[height=0.13\textheight]{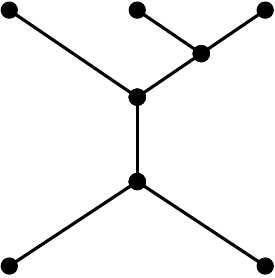}}}
  \qquad
  {\raisebox{0\textheight}
  {\includegraphics[height=0.13\textheight]{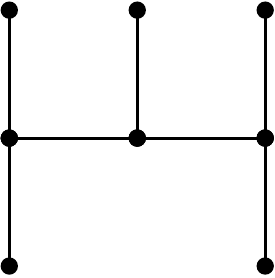}}}
  \\
  \vspace{0.05\textheight}
  {\raisebox{0\textheight}
  {\includegraphics[height=0.095\textheight]{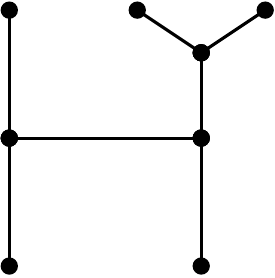}}}
  \qquad
  {\raisebox{0\textheight}
  {\includegraphics[height=0.095\textheight]{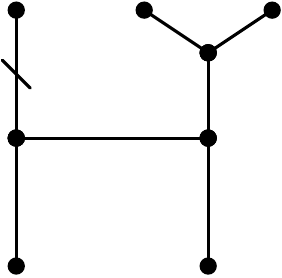}}}
  \qquad
  {\raisebox{0\textheight}
  {\includegraphics[height=0.095\textheight]{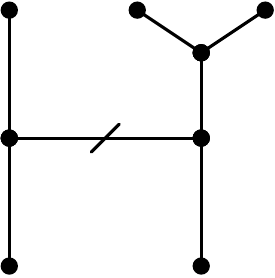}}}
  \qquad
  {\raisebox{0\textheight}
  {\includegraphics[height=0.095\textheight]{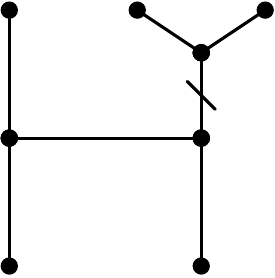}}}
  \qquad
  {\raisebox{0\textheight}
  {\includegraphics[height=0.095\textheight]{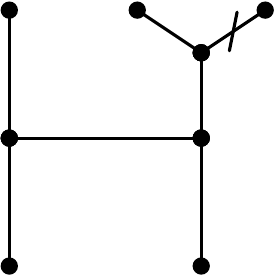}}}
  \\
  \vspace{0.03\textheight}
  {\raisebox{0\textheight}
  {\includegraphics[height=0.095\textheight]{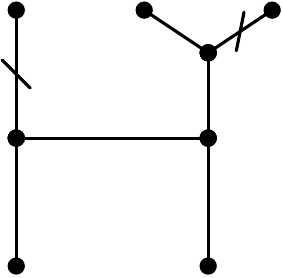}}}
  \qquad
  {\raisebox{0\textheight}
  {\includegraphics[height=0.095\textheight]{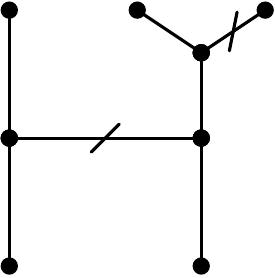}}}
  \qquad
  {\raisebox{0\textheight}
  {\includegraphics[height=0.095\textheight]{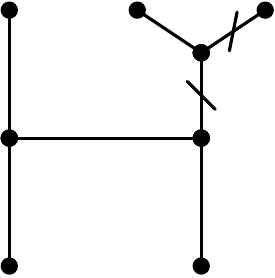}}}
  \qquad
  {\raisebox{0\textheight}
  {\includegraphics[height=0.095\textheight]{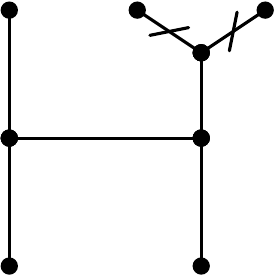}}}
  \caption{(a) The topologies relevant to 2-to-3 scattering; (b) The
  graphs corresponding to the first topology in (a).}
  \label{fig:2to3}
\end{figure}

The correspondence also works out for $2 \to 3$ at tree-level. In this
case, there are three topologies to consider, as illustrated in figure
\ref{fig:2to3}(a).   Figure   \ref{fig:2to3}(b)   shows   the   graphs
corresponding to the first  topology in figure \ref{fig:2to3}(a). Each
graph should be  summed over all allowed time  orderings (i.e.~subject
to the rule that there must always be at least one retarded propagator
heading  forwards from  any vertex)  and agreement  with  the S-matrix
calculation follows,  e.g.~for the graphs  in figure \ref{fig:2to3}(b)
the relative weights  are (in order from the first  to the last graph)
$3-1-1-1-6+2+2+2+1  = 1$  and we  have doubled  the  contribution from
graphs $5-8$  to account  for the contribution  where the  Feynman and
retarded  propagators are swapped  on the  rightmost pair  of outgoing
legs.  The two other types  of graph shown in figure \ref{fig:2to3}(a)
also each have a total weight equal to $1$. An all orders proof of the
equivalence  with  the  S-matrix,  in  the  case  of  positive  energy
plane-wave scattering, is provided in section~\ref{sec:smatrix}.

\section{Path integral representation}
\label{sec:pathintegral}

In this section  we will explain the connection  of the operator-level
amplitudes  in the  previous  section to  the  retarded amplitudes  of
thermal  field theory. To  this end,  we will  apply the  ``in-in'' or
closed-time  path  (CTP)  formalism   due  to  Schwinger  and  Keldysh
\cite{Schwinger:1960qe, Keldysh:1964ud}.  In real-time formulations of
quantum  field   theory  at   finite  temperature  and   density  (see
e.g.~\cite{Millington:2012pf}  and references  therein),  it is  known
that physical  reaction rates must  be calculated from  the absorptive
parts of {\it{retarded}} self-energies  in order to obtain the correct
quantum statistics  \cite{Kobes:1990ua, vanEijck:1992mq}. We emphasise
that  retarded self-energies  have arisen  naturally in  our treatment
thus far (e.g.~see the diagrams in figure \ref{fig:f3x}).

The  starting  point of  the  ``in-in"  generating  functional is  the
partition function:
\begin{equation}
  \label{eq::partition}
  Z\ =\ \mathrm{tr}\,\rho\;
\end{equation}
where the density operator  $\rho$ represents the detection subsystem,
which we suppose to act  in the time interval $[T_1,\ T_2]$. Contained
in this generating functional  is the amplitude $\Gamma_{JK}$ from the
previous  section, since,  in  the Heisenberg  picture,  we can  write
$\Gamma_{JK} = \langle 0 | \rho | 0 \rangle$ where
\beq
  \rho = \text{T}\; \text{exp} \left( i \int_T^\infty \dd^4 x \; 
  K_x \phi_x^\text{H} \right)~.
\eeq 

To  build the  path  integral, we  prepare  a set  of $N$  independent
detectors,  each described by  a density  operator $\rho_i$  and whose
actions  have exclusive support  over infinitesimal  intervals $\Delta
t\: =\:  t_{i+1}\:-\:t_i\:=\:(T_1\:-\:T_2)/N$, where $t_i\:\in\:[T_1,\
T_2]$. In the continuous limit, $N\:\to\:\infty$, the density operator
of  the combined system  of detectors  may be  written as  the product
integral
\begin{equation}
  \rho\ =\ \prod_{T_1}^{T_2}\rho(t)^{\mathrm{d}t}\;.
\end{equation}
In order  to generate a path-integral representation  of the partition
function $Z$,  we imagine  perturbing the evolution  of the  system by
means  of some  unphysical test  source  $j_x$. Thus,  we insert  into
eq.~(\ref{eq::partition}) unity in the form
\begin{equation}
  \overline{\mathrm{T}}\exp\bigg[-i\int_{T_A}^{T_B}\mathrm{d}^4x\;j_x \,
  \phi^{\mathrm{H}}_x\bigg]\mathrm{T}\exp\bigg[
  i\int_{T_A}^{T_B}\mathrm{d}^4x\;j_x
  \, \phi^{\mathrm{H}}_x\bigg]\ =\ \mathbb{I}\;.
\end{equation}
Notice  that this  insertion  will generate  two  paths of  evolution:
$\mathcal{C}_+$,  running forwards  in time  from $T_A$  to  $T_B$ and
$\mathcal{C}_-$,  running backwards from  $T_B$ and  $T_A$. It  is the
presence of  these two  anti-parallel integration contours  that gives
rise  to  the  closed-time path  $\mathcal{C}\:=\:\mathcal{C}_+\:\cup\
\mathcal{C}_-$ of the ``in-in" formalism. Hereafter, objects with time
arguments confined  to the positive (time-ordered)  branch are denoted
by a  subscript `$+$'  and those with  time arguments confined  to the
negative  (anti-time-ordered)  branch   are  denoted  by  a  subscript
`$-$'.  We note  that  the time  $T_A$  is the  boundary  time of  the
evolution  of   the  system  at  which  the   initial  conditions  are
specified.\footnote{For a discussion of the importance of keeping track
of   this    boundary   time   in    non-equilibrium   phenomena   see
\cite{Millington:2012pf}.}   Ultimately,  we   will  take  the  limits
$T_A\:\to\:-\:\infty$ assuming that, asymptotically, the system is the
free vacuum.

By   further   inserting  complete   sets   of   eigenstates  of   the
Heisenberg-picture   field  operator  $\phi^{\mathrm{H}}_x$,   we  may
develop the path-integral representation of the generating functional:
\begin{align}
  Z[j^a]\ &=\ \int[\mathrm{d}\phi^a_{\mathbf{z}}]
  \braket{\phi^-_{\mathbf{z}},T_B|\rho|\phi^+_{\mathbf{z}},T_B}
  \exp\bigg[i\bigg(S[\phi^a;T_A,T_B]\:
  +\:i\int_{T_A}^{T_B}\!\mathrm{d}^4x\;\eta_{ab}j^a_x\phi^b_x\bigg)\bigg]\;,
\end{align}
where                $[\mathrm{d}\phi^a_{\mathbf{z}}]                =
\prod_{T_A}^{T_B}[\mathrm{d}\phi^a_t(\mathbf{z})]^{\mathrm{d}t}$
denotes   functional   integration   over   `$+$'  and   `$-$'   field
configurations. The test  sources and fields have been  written in the
doublet      notation      employed     in      \cite{Calzetta:1986cq,
Calzetta:1986ey,Jordan:1986ug}, where
\begin{subequations}
  \begin{align}
    j^a_x\ &=\ \Big(j^+_x,\ j^-_x \Big)\;,\qquad &
    j_{a,\,x}\ &=\ \eta_{ab}j^b_x\ =\ \Big(j^+_x,\ -\:j^-_x\Big)\\
    \phi^a_x\ &=\ \Big(\phi^+_x,\ \phi^-_x\Big)\;,\qquad &
    \phi_{a,\,x}\ &=\ \eta_{ab}\phi^b_x\ =\ \Big(\phi^+_x,\ -\:\phi^-_x\Big)
  \end{align}
\end{subequations}
and  $\eta_{ab}\:=\:\mathrm{diag}(1,\ -1)$.   Hereafter,  CTP indices,
labelling  the confinement  of objects  to the  positive  and negative
branches  of the  CTP contour,  are  denoted by  the lower-case  Roman
characters  $a,b\:=\:1,2\:\equiv\:+,-$.  In  the  same  notation,  the
action $S[\phi^a;T_A,T_B]$ may be written
\begin{equation}
  S[\phi^a;T_A,T_B]\ =\ \int_{T_A}^{T_B}\!\mathrm{d}^4 x\;
  \bigg[\frac{1}{2}\eta_{ab}\Big(\partial_{\mu}\phi^a_x\partial^{\mu}\phi^b_x\:
  -\:m^2\phi^a_x\phi^b_x\Big)\:+\:\mathcal{L}^{\mathrm{int}}(\phi^a)\bigg]\;,
\end{equation}
where the interaction part is
\begin{align}
  \label{eq::Lint}
  \mathcal{L}^{\mathrm{int}}(\phi^a)\ &=\ \eta_{ab}J^a_x\phi^b_x\:
  -\:\frac{g}{3!}\eta_{abc}\phi_x^a\phi_x^b\phi_x^c\;.
\end{align}
It contains the physical emission sources 
\beq
  J^a = (J_x,\ J_x )~.
\eeq
The tensor $\eta_{abc}$, appearing in eq.~(\ref{eq::Lint}), is defined such that
\begin{equation}
  \eta_{abc}\ =\ \begin{cases} +1\;,\qquad &
    a\:=\:b\:=\:c\:=\:1\\ -1\;,\qquad &
    a\:=\:b\:=\:c\:=\:2\\ 0\;,\qquad &
    \mathrm{otherwise}.
  \end{cases}
\end{equation}

We may introduce an operator $\sqrt{\rho}$ and write the kernel of the
density operator in the form
\begin{equation}
  \braket{\phi^-_{\mathbf{z}},T_B|\rho|\phi^+_{\mathbf{z}},T_B}\ =\
  \braket{\phi^-_{\mathbf{z}},T_B|(\sqrt{\rho})^2|\phi^+_{\mathbf{z}},T_B}\;.
\end{equation}
Again,  by inserting complete  sets of  eigenstates of  the Heisenberg
field operator, we obtain
\begin{align}
  \braket{\phi^-_{\mathbf{z}},T_B|\rho|\phi^+_{\mathbf{z}},T_B}\ &\sim \
  \int[\mathrm{d}\phi^a_{T_2}(\mathbf{z})]\;\braket{\phi^-_{\mathbf{z}},T_1|
    \sqrt{\rho}|\phi^-_{\mathbf{z}},T_2}
  \:\braket{\phi^+_{\mathbf{z}},T_2|\sqrt{\rho}|\phi^+_{\mathbf{z}},T_1}
  \nonumber\\
  &\sim\ \int[\mathrm{d}\phi^a_{\mathbf{z}}]\;
  \exp\Big(iK[\phi^a;T_1,T_2]\Big)\;,
\end{align}
where       $[\mathrm{d}\phi_{\mathbf{z}}^a]       \:       =       \:
\prod_{T_1}^{T_2}[\mathrm{d}\phi_t^a(\mathbf{z})]^{\mathrm{d}t}$    and
to  simplify  matters we  henceforth  assume $T_1  =  T_B  = T$  (this
definition of $T$ matches that in the previous section).

In general,  the exponent $K[\phi^a;T_1,T_2]$ will be  expressed as an
infinite series  of convolutions of  poly-local sources and  fields of
the                                                                form
$K_{abc\dots,xyz\dots}\phi^a_x\phi^b_y\phi^c_z\cdots$.   However,  for
our  purposes, we  shall take  $K[\phi^a;T_1,T_2]$ to  contain  only a
local source, i.e.
\begin{equation}
  \label{eq::kerneldef}
  K[\phi^a;T_1,T_2]\ =\ \frac{1}{2}\int_{T_1}^{T_2}\!\mathrm{d}^4x\;
  K_{a,\,x} \; \phi^{a}_{x}\;,
\end{equation}
where
\begin{equation}
  K^a_x\ =\ \Big(K_x,\ -K_x\Big)~.
\end{equation}
Notice that $K^a_x$  differs by a sign in  the second element relative
to the  emission source  $J^a_x$. This relative  sign and  the overall
factor of $1/2$ in eq.~(\ref{eq::kerneldef}) arise from writing $\rho\
=\ (\sqrt{\rho})^2$.

\begin{figure}[t]
\centering
$\begin{array}{c}
  \includegraphics[width=100mm]{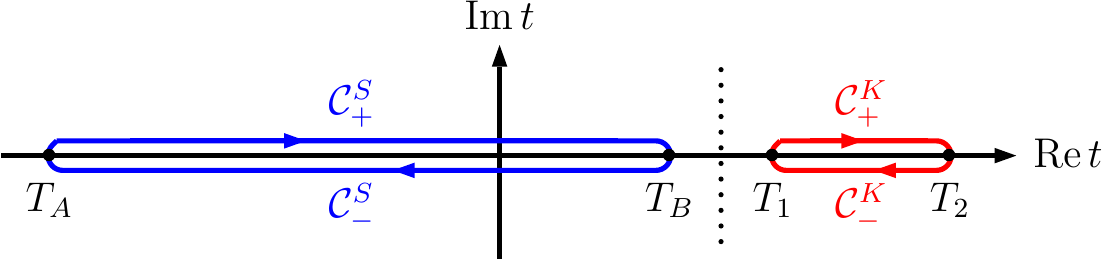}
\end{array}$
  \caption{The two  closed contours corresponding to  the time
  integrals in eq.~(\ref{eq::ininstart}).\label{fig:contour}}
\end{figure}

Thus, we arrive at the  form of the ``in-in" generating functional for
our choice of density operator:
\begin{align}
  \label{eq::ininfield}
  Z[j^a]\ &=\ \int[\mathrm{d}\phi^a_{\mathbf{z}}]
  \exp\bigg[\frac{i}{2}\int_{T_1}^{T_2}\!\mathrm{d}^4x\;
  \eta_{ab}K^a_x\phi^b_x\bigg]
  \exp\bigg[i\bigg(S[\phi^a;T_A,T_B]\:
  +\:\int_{T_A}^{T_B}\!\mathrm{d}^4x\;
  \eta_{ab}j^a_x\phi^b_x\bigg)\bigg]\;.
\end{align}
Completing the square in the free part of the action, we write
\begin{equation}
  \phi'{}^a_x\ =\ \phi^a_x\:
  -\:i\int_{T_A}^{T_B}\!\mathrm{d}^4y\;\Delta^{ab}_{xy}\,j_{b,\, y}\;,
\end{equation}
where
\begin{equation}
  \Delta^{ab}_{xy}\ =\ \begin{bmatrix}
    \Delta^{\mathrm{F}}_{xy} & \Delta^{<}_{xy} \\
    \Delta^{>}_{xy} & \Delta^{\mathrm{D}}_{xy}
  \end{bmatrix}
\end{equation}
is the free  $2\:\times\:2$ CTP matrix propagator. With  this shift in
the field,  we may  recast the ``in-in"  generating functional  in the
form
\begin{align}
  \label{eq::ininstart}
  Z[j^a]\ &=\ Z_0[0]\exp\bigg[\frac{1}{2}\int_{T_1}^{T_2}\!\mathrm{d}^4
  x\;K_{a,\,x}\delta^a_x\bigg]
  \exp\bigg[i\int_{T_A}^{T_B}\!\mathrm{d}^4x\;\mathcal{L}^{\mathrm{int}}
  \big(-i\delta^a_x\big)\bigg]
  \nonumber\\&\qquad \qquad \qquad \qquad \qquad \qquad \qquad
  \times\exp\bigg[
  -\frac{1}{2}\iint_{-\infty}^{+\infty}\!\mathrm{d}^4x\,\mathrm{d}^4y\;
  j_{a,\,x}\,\Delta^{ab}_{xy}\,j_{b,\,y}\bigg] 
\end{align}
where  $Z_0[0]$  is  the  generating  functional  in  the  absence  of
interactions and  for vanishing test  sources $j_{a,\,x}$ and  we have
introduced the short-hand notation
\begin{equation}
  \delta^a_x\ \equiv\ \frac{\delta}{\delta j_{a,\,x}}
\end{equation}
for functional derivatives  with respect to the test  sources. The two
time integrals over the intervals $[T_A,\ T_B]$ and $[T_1,\ T_2]$ have
given  rise  to  two  closed-time  paths,  as  illustrated  in  figure
\ref{fig:contour}, each  corresponding to  one of the  subsystems into
which the system has been partitioned. We will take $T_A \to -\infty$,
$T_B = T_1 = T$ and $T_2 \to +\infty$.

By means of an orthogonal transformation, we may rotate to the Keldysh
basis (e.g.~see \cite{Millington:2012pf,vanEijck:1994rw}):
\begin{equation}
  \widetilde{\Delta}^{ab}_{xy}\ =\ O^a_{\ c}O^b_{\ d}\Delta^{cd}_{xy}\ 
  =\ \begin{bmatrix}
    0 & \Delta^{\mathrm{A}}_{xy} \\
    \Delta^{\mathrm{R}}_{xy} & \Delta^{1}_{xy}
  \end{bmatrix}\;,\qquad
  O^{ab}\ =\ \frac{1}{\sqrt{2}}
  \begin{bmatrix}
    1 & 1 \\
    1 & -1
  \end{bmatrix}
\end{equation}
in which  the elements  of the CTP  propagator comprise  the retarded,
advanced and Hadamard propagators (see appendix \ref{app:props}). With
this transformation
\begin{align}
  \label{eq::iningen1}
  Z[\widetilde{j}^a]\ &=\
  Z_0[0]\exp\bigg[\frac{1}{\sqrt{2}}\int_{T_1}^{T_2}\!\mathrm{d}^4
  x\;K_x\frac{\delta}{\delta
  \widetilde{j}_{-,\,x}}\bigg]\exp\bigg[i\int_{T_A}^{T_B}\!\mathrm{d}^4x\;
  \mathcal{L}^{\mathrm{int}}\bigg(\frac{1}{i}\frac{\delta}{\delta
  \widetilde{j}_{a,\,x}}\bigg)\bigg]
  \nonumber\\&\qquad
  \times\:\exp\bigg[-\frac{1}{2}\iint_{-\infty}^{+\infty}\!\mathrm{d}^4x\,
  \mathrm{d}^4y\;\widetilde{j}_{a,\,x}\widetilde{\Delta}^{ab}_{xy}\,
  \widetilde{j}_{b,\,y}\bigg]\;,
\end{align}
where we have defined
\begin{equation}
  \widetilde{j}^a_x\ =\ \Big(\widetilde{j}^+_x,\ \widetilde{j}^-_x\Big)
\end{equation}
with
\begin{equation}
  \widetilde{j}^{\pm}_x\ =\ \frac{1}{\sqrt{2}}\Big(j^+_x\:\pm\:j^-_x\Big)\;.
\end{equation}
Subsequently,  contracting  the  CTP   indices  in  the  exponents  of
eq.~(\ref{eq::iningen1}), we obtain
\begin{align}
  \label{eq::iningen2}
  Z[\widetilde{j}^a]\ &=\
  Z_0[0]\exp\bigg[\frac{1}{\sqrt{2}}\int_{T_1}^{T_2}\!\mathrm{d}^4
  x\;K_x\widetilde{\delta}^-_x\bigg]\exp\bigg[i\int_{T_A}^{T_B}\!\mathrm{d}^4x\;
  \mathcal{L}^{\mathrm{int}}\big(-i
  \delta^{a}_{x}\big)\bigg]
  \nonumber\\&\quad
  \times\exp\bigg[-\frac{1}{2}\iint_{-\infty}^{+\infty}\!\mathrm{d}^4x\,
  \mathrm{d}^4y\;\Big(\widetilde{j}^-_{x}\,\Delta^{\mathrm{R}}_{xy}\,
  \widetilde{j}^+_{y}\:+\:\widetilde{j}^+_{x}\,\Delta^{\mathrm{A}}_{xy}\,
  \widetilde{j}^-_{y}\:+\:\widetilde{j}^{-}_x\,\Delta^{1}_{xy}\,
  \widetilde{j}^-_{y}\Big) \bigg]
\end{align}
in which
\begin{equation}
  \mathcal{L}^{\mathrm{int}}\big(-i\delta^{a}_{x}\big)\ =\
  -\:i\bigg[\sqrt{2}J_x\widetilde{\delta}^+_x\:
  +\:\frac{g}{3!\sqrt{2}}\bigg(\big(\widetilde{\delta}^+_x\big)^3\:
  +\:3\widetilde{\delta}^+_x\big(\widetilde{\delta}^-_x\big)^2\bigg)\bigg]\;
\end{equation}
and we have used the fact that
\begin{equation}
  \delta^{\pm}_{x}\ =\ \frac{1}{\sqrt{2}}\big(\widetilde{\delta}^+_x\:
  \pm\:\widetilde{\delta}^-_x\big)\;.
\end{equation}
After suitable changes of variables and using the relations
\begin{equation}
  \Delta^{\mathrm{R}}_{xy}\ =\ \Delta^{\mathrm{A}}_{yx}\;,\qquad 
  \Delta^{1}_{xy}\ =\ 2\Delta^{\mathrm{F}}_{xy}\:-\:\Delta^{\mathrm{R}}_{xy}\:
  -\:\Delta^{\mathrm{A}}_{xy}\;,
\end{equation}
the  Keldysh  representation  of  the ``in-in"  generating  functional
eq.~(\ref{eq::iningen2}) may re-expressed as
\begin{align}
  \label{eq::iningenfin}
  Z[\widetilde{j}^a]\ &=\
  Z_0[0]\exp\bigg[\frac{1}{\sqrt{2}}\int_{x}K_x\,\widetilde{\delta}^{-}_x\bigg]
  \exp\bigg[\sqrt{2}\int_xJ_x\,\widetilde{\delta}^+_x\bigg]
  \exp\bigg\{\frac{1}{3!\sqrt{2}}
  \int_{x}\Big[\big(\widetilde{\delta}^+_x\big)^3\:
  +\:3\big(\widetilde{\delta}^-_x\big)^2\widetilde{\delta}^+_x\Big]\bigg\}
  \nonumber\\&\qquad
  \times\:\exp\bigg\{-\int_{xy}\Big[
  \widetilde{j}^-_{x}\,\Delta^{\mathrm{R}}_{xy}\,\widetilde{j}^+_{y}\:
  +\:\widetilde{j}^-_{x}\Big(\Delta^{\mathrm{F}}_{xy}\:
  -\:\Delta^{\mathrm{R}}_{xy}\Big)\widetilde{j}^-_{y}\Big] \bigg\}\;.
\end{align}
Equation   (\ref{eq::iningenfin})   is  the   main   result  of   this
section.  Notice in  particular  that {\it  {the  physical source  $J$
couples only to the retarded propagator}}.

\paragraph{An example.}  By way of  illustration, we now  consider the
specific case  of the one-to-two amplitude, obtained  by expanding the
exponentials  in  eq.~(\ref{eq::iningenfin}) to  second  order in  the
detection sources $K$ and to first order in the emission sources $J$:
\begin{align}
  \Gamma^{1\to2}_{JK}\ &=\frac{g}{8}\int_{xyzw}\;K_xK_yJ_w
  \widetilde{\delta}^{-}_{x}\widetilde{\delta}^{-}_{y}
  (\widetilde{\delta}^{-}_{z})^2\widetilde{\delta}^{+}_{z}
  \widetilde{\delta}^{+}_{w}
  \nonumber\\&\qquad \times \exp\bigg\{-\int_{x'y'}
  \Big[\,\widetilde{j}^-_{x'}\,\Delta^{\mathrm{R}}_{x'y'}\,
  \widetilde{j}^+_{y'}\:
  +\:\widetilde{j}^-_{x'}\Big(\Delta^{\mathrm{F}}_{x'y'}\:
  -\:\Delta^{\mathrm{R}}_{x'y'}\Big)\widetilde{j}^-_{y'}\Big]
  \bigg\}\bigg|_{\widetilde{j}^a\:=\:0}\;.
\end{align}
Note that this amplitude is obtained directly from $Z[\tilde{j}^a]$.
Acting with the rightmost functional derivative, we have
\begin{align}
  \Gamma^{1\to2}_{JK}\ &= - \frac{g}{8}\int_{xyzw1}\;K_xK_yJ_w
  \widetilde{\delta}^{-}_{x}\widetilde{\delta}^{-}_{y}
  (\widetilde{\delta}^{-}_{z})^2\widetilde{\delta}^{+}_{z}
  \widetilde{j}^-_{1}\Delta^{\mathrm{R}}_{1w}
  \nonumber\\&\qquad
  \times\exp\bigg\{-\int_{x'y'}\Big[\,
  \widetilde{j}^-_{x'}\,\Delta^{\mathrm{R}}_{x'y'}\,\widetilde{j}^+_{y'}\:
  +\:\widetilde{j}^-_{x'}\Big(\Delta^{\mathrm{F}}_{x'y'}\:
  -\:\Delta^{\mathrm{R}}_{x'y'}\Big)\widetilde{j}^-_{y'}\Big]
  \bigg\}\bigg|_{\widetilde{j}^a\:=\:0}\;.
\end{align}
Performing the  remaining $\widetilde{j}^+$ functional  derivative, we
may neglect the first term in the exponent, giving
\begin{align}
  \Gamma^{1\to2}_{JK}\ &=\frac{g}{8}\int_{xyzw12}K_xK_yJ_w
  \widetilde{\delta}^{-}_{x}\widetilde{\delta}^{-}_{y}
  (\widetilde{\delta}^{-}_{z})^2
  \nonumber\\&\qquad \qquad
  \times \widetilde{j}^-_{1}\widetilde{j}^-_{2}\Delta^{\mathrm{R}}_{2z}
  \Delta^{\mathrm{R}}_{1w}\exp\bigg\{-\int_{x'y'}\widetilde{j}^-_{x'}
  \Big(\Delta^{\mathrm{F}}_{x'y'}\:-\:\Delta^{\mathrm{R}}_{x'y'}\Big)
  \widetilde{j}^-_{y'} \bigg\}\bigg|_{\widetilde{j}^-\:=\:0}\;.
\end{align}
Differentiating     again     and      using     the     fact     that
$\Delta^{\mathrm{R}}_{xx}\:=\:0$, we obtain
\begin{align}
  \Gamma^{1\to2}_{JK}\ &=\frac{g}{8}\int_{xyzw2}K_xK_yJ_w
  \widetilde{\delta}^{-}_{x}\widetilde{\delta}^{-}_{y}
  \widetilde{\delta}^{-}_{z}j^-_{2}\Delta^{\mathrm{R}}_{2z}
  \bigg[\Delta^{\mathrm{R}}_{zw}\:-\:\int_{13}\widetilde{j}^-_{1}
  \widetilde{j}^-_{3}\Delta^{\mathrm{R}}_{1w}
  \Big(2\Delta^{\mathrm{F}}_{3z}\:-\:\Delta^{\mathrm{R}}_{3z}\:
  -\:\Delta^{\mathrm{A}}_{3z}\Big)\bigg]
  \nonumber\\&\qquad
  \times\:\exp\bigg\{-\int_{x'y'}\widetilde{j}^-_{x'}
  \Big(\Delta^{\mathrm{F}}_{x'y'}\:-\:\Delta^{\mathrm{R}}_{x'y'}\Big)
  \widetilde{j}^-_{y'} \bigg\}\bigg|_{\widetilde{j}^-\:=\:0}\;.
\end{align}
Finally, keeping only the connected diagrams, we have
\begin{align}
  \label{eq::withA}
  \Gamma^{1\to2}_{JK}\ =-\frac{g}{4}\int_{xyzw12}K_xK_yJ_w
  \widetilde{\delta}^{-}_{x}\widetilde{\delta}^{-}_{y}
  \widetilde{j}^-_{1}\widetilde{j}^-_{2}\Delta^{\mathrm{R}}_{1z}
  \Big(2\Delta^{\mathrm{F}}_{2z}\:-\:\Delta^{\mathrm{R}}_{2z}\:
  -\:\Delta^{\mathrm{A}}_{2z}\Big)\Delta^{\mathrm{R}}_{zw}\;,
\end{align}
where  integration  variables  have  been  relabelled  for  notational
convenience.  Performing  the  remaining  functional  derivatives,  we
arrive at the result
\begin{equation}
  \Gamma^{1\to2}_{JK}\ =-\frac{g}{2}\int_{xyzw}K_xK_y\;
  \Delta^{\mathrm{R}}_{xz}\Big(2\Delta^{\mathrm{F}}_{yz}\:
  -\:\Delta^{\mathrm{R}}_{yz}\:
  -\:\Delta^{\mathrm{A}}_{yz}\Big)\Delta^{\mathrm{R}}_{zw}\,J_w\;,
  \label{eq:1to2p}
\end{equation}
which is in agreement with eq.~(\ref{eq:1to2}) since $x^0,y^0 > T$ and
$z^0 < T$, so the advanced contribution vanishes.

\section{Relation to S-matrix}
\label{sec:smatrix}

In this  section, we will show  that the amplitudes  calculated in the
preceding  sections  are  equivalent  to  the  corresponding  S-matrix
amplitudes  for the  scattering  of positive  energy plane-waves.   We
begin by  considering a  general set of  tree-level graphs  with $N_K$
outgoing external legs and $N_J$ incoming external legs. The number of
vertices $V$ and propagators $P$ are given by
\begin{equation}
  \label{eq:V}
  V\ =\ N\:-\:2\;,\qquad 
  P\ =\ 2N\:-\:3
\end{equation}
in  which $N\:=\:N_K\:+\:N_J$ is  the total  number of  external legs.
Expanding  each of the  exponentials in  eq.~(\ref{eq::iningenfin}) to
the appropriate order, we have
\begin{align}
  \label{eq:gengamma}
  \Gamma^{N_J\to N_K}_{JK}\ &=\ (-1)^P\bigg(\frac{g}{\sqrt{2}}\bigg)^{V}
  \frac{1}{V^+!}\bigg(\frac{1}{3!}\int_x
  \big(\widetilde{\delta}^+_x\big)^3\bigg)^{V^+}
  \frac{1}{V^-!}\bigg(\frac{1}{2!}\int_x
  \big(\widetilde{\delta}^-_x\big)^2\widetilde{\delta}^+_x\bigg)^{V^-}
  \nonumber\\&\qquad
  \times\frac{1}{N_K!}\bigg(\frac{1}{\sqrt{2}}\int_x
  K_x\,\widetilde{\delta}^-_x\bigg)^{N_K}
  \frac{1}{N_J!}\bigg(\sqrt{2}\int_xJ_x\,\widetilde{\delta}^+_x\bigg)^{N_J}
  \nonumber\\&\qquad
  \times\frac{1}{P^+!}\bigg(\int_{xy}\widetilde{j}^-_{x}\,
  \Delta^{\mathrm{R}}_{xy}\,\widetilde{j}^+_{y}\bigg)^{P^+}
  \frac{1}{P^-!}\bigg(\frac{1}{2}\int_{xy}\widetilde{j}^-_{x}
  \Big(2\Delta^{\mathrm{F}}_{xy}\:-\:\Delta^{\mathrm{R}}_{xy}\:
  -\:\Delta^{\mathrm{A}}_{xy}\Big)\widetilde{j}^-_{y}\bigg)^{P^-}\;,
\end{align}
where $V\:=\:V^+\:+\:V^-$ and  $P\:=\:P^+\:+\:P^-$. For convenience of
notation, we have left implicit  the fact that the $\widetilde{j}$ are
set to zero externally.

Using    eq.~(\ref{eq:V}),    the    factors    of    $\sqrt{2}$    in
eq.~(\ref{eq:gengamma}) can be combined to give
\begin{align}
  \Gamma^{N_J\to N_K}_{JK}\ &=\ (-1)^P2^{1-N_K}g^V\frac{1}{V^+!}
  \bigg(\frac{1}{3!}\int_x\big(\widetilde{\delta}^+_x\big)^3\bigg)^{V^+}
  \frac{1}{V^-!}\bigg(\frac{1}{2!}\int_x
  \big(\widetilde{\delta}^-_x\big)^2\widetilde{\delta}^+_x\bigg)^{V^-}
  \nonumber\\&\qquad
  \times\frac{1}{N_K!}\bigg(\int_xK_x\,\widetilde{\delta}^-_x\bigg)^{N_K}
  \frac{1}{N_J!}\bigg(\int_xJ_x\,\widetilde{\delta}^+_x\bigg)^{N_J}
  \nonumber\\&\qquad
  \times\frac{1}{P^+!}\bigg(\int_{xy}\widetilde{j}^-_{x}\,
  \Delta^{\mathrm{R}}_{xy}\,\widetilde{j}^+_{y}\bigg)^{P^+}
  \frac{1}{P^-!}\bigg(\frac{1}{2}\int_{xy}\widetilde{j}^-_{x}
  \Big(2\Delta^{\mathrm{F}}_{xy}\:-\:\Delta^{\mathrm{R}}_{xy}\:
  -\:\Delta^{\mathrm{A}}_{xy}\Big)\widetilde{j}^-_{y}\bigg)^{P^-}\;.
\end{align}
Performing the functional derivatives in the sources, we obtain
\begin{align}
  \label{eq:gengamma2}
  \Gamma^{N_J\to N_K}_{JK}\ &=\ (-1)^P2^{1-N_K}g^V\frac{1}{V^+!}
  \bigg(\frac{1}{3!}\int_x\big(\widetilde{\delta}^+_x\big)^3\bigg)^{V^+}
  \frac{1}{V^-!}\bigg(\frac{1}{2!}\int_x\big(\widetilde{\delta}^-_x\big)^2
  \widetilde{\delta}^+_x\bigg)^{V^-}
  \nonumber\\&\qquad
  \times\frac{1}{N_K!}\bigg(\int_{xy}K_x\,\Delta^{\mathrm{R}}_{xy}\,
  \widetilde{j}^+_y\bigg)^{N^{+}_K}\bigg(\int_{xy}K_x
  \Big(2\Delta^{\mathrm{F}}_{xy}\:-\:\Delta^{\mathrm{R}}_{xy}\Big)
  \widetilde{j}^-_y\bigg)^{N^{-}_K}
  \nonumber\\&\qquad
  \times\frac{1}{N_J!}\bigg(\int_{xy}
  \widetilde{j}^-_x\,\Delta^{\mathrm{R}}_{xy}\,J_y\bigg)^{N_J}
  \frac{1}{(P^+-N_K^{+}-N_J)!}\bigg(\int_{xy}\widetilde{j}^-_{x}\,
  \Delta^{\mathrm{R}}_{xy}\,\widetilde{j}^+_{y}\bigg)^{P^+-N_K^{+}-N_J}
  \nonumber\\&\qquad
  \times\frac{1}{(P^--N_K^{-})!}\bigg(\frac{1}{2}\int_{xy}
  \widetilde{j}^-_{x}\Big(2\Delta^{\mathrm{F}}_{xy}\:
  -\:\Delta^{\mathrm{R}}_{xy}\:-\:\Delta^{\mathrm{A}}_{xy}\Big)
  \widetilde{j}^-_{y}\bigg)^{P^--N_K^-}\;,
\end{align}
where  $N_K\: =\: N_{K}^{+}\:+\:N_{K}^{-}$.  Notice that  the advanced
contribution does not appear in the outgoing external legs since $K_x$
acts only in the future of all other vertices.

If all of  the external four-momenta are on-shell  the purely on-shell
combination
$2\Delta^{\mathrm{F}}_{xy}\:-\:\Delta^{\mathrm{R}}_{xy}\:-\:\Delta^{\mathrm{A}}_{xy}$
cannot occur in  the internal lines of tree-level  graphs by virtue of
energy-momentum      conservation.      As      such,      we      set
$P^-\:=\:N_K^{-}$. Equation  (\ref{eq:gengamma2}) then reduces  to the
following:
\begin{eqnarray}
  \label{eq:gengamma3}
  \Gamma^{N_J\to N_K}_{JK}\ &=&\
  (-1)^P2^{1-N_K}g^V\frac{1}{V^+!}\bigg(\frac{1}{3!}\int_x
  \big(\widetilde{\delta}^+_x\big)^3\bigg)^{V^+}
  \frac{1}{V^-!}\bigg(\frac{1}{2!}\int_x\big(\widetilde{\delta}^-_x\big)^2
  \widetilde{\delta}^+_x\bigg)^{V^-}
  \nonumber\\&
  \times&\frac{1}{N_K!}\bigg(\int_{xy}K_x\,\Delta^{\mathrm{R}}_{xy}\,
  \widetilde{j}^+_y\bigg)^{N^{+}_K}\bigg(\int_{xy}K_x
  \Big(2\Delta^{\mathrm{F}}_{xy}\:-\:\Delta^{\mathrm{R}}_{xy}\Big)
  \widetilde{j}^-_y\bigg)^{N^{-}_K}
  \nonumber\\&
  \times&\frac{1}{N_J!}\bigg(\int_{xy}\widetilde{j}^-_x\,
  \Delta^{\mathrm{R}}_{xy}\,J_y\bigg)^{N_J}\frac{1}{(P-N)!}
  \bigg(\int_{xy}\widetilde{j}^-_{x}\,\Delta^{\mathrm{R}}_{xy}\,
  \widetilde{j}^+_{y}\bigg)^{P-N}\;.
\end{eqnarray}
This is to be compared with the corresponding term in the expansion of
the usual formula for the S-matrix:
\begin{equation}
  S\ =\ :\exp\bigg[\int_x\;\phi^{\mathrm{in}}_x(\Box_x^2+m^2)\delta_x\bigg]:
  \exp\bigg[\int_x\frac{g}{3!}\delta_x^3\bigg]
  \exp\bigg[-\frac{1}{2}\int_{x,y} \; j_x\,
  \Delta^\text{F}_{xy}\,j_y\bigg]\bigg|_{j\:=\:0}\;.
  \label{eq:LSZ}
\end{equation}
As in eq.~(\ref{eq:LSZ}), we can  affect the LSZ reduction to map from
the vacuum amplitude to the S-matrix by promoting the external sources
to operators in Fock space, i.e.
\bea
  K_x & \to & \phi^{\mathrm{out}}_x \; (\Box_x^2\:+\:m^2)\;,
  \nonumber \\
  J_x\ & \to &\phi^{\mathrm{in}}_x \;  (\Box_x^2\:+\:m^2)\;~.
\eea
On  contraction  with the  $N_K$-particle  ``out'' and  $N_J$-particle
``in'' Fock  states, $\Gamma_{JK}^{N_J \to  N_K}$ gives rise to  a sum
over  all  possible  connected   topologies.  In  the  usual  S-matrix
approach, we  would obtain  a single graph  for each topology.  In the
case of eq.~(\ref{eq:gengamma3}) however, for each topology, we obtain
a set  of graphs  with each graph  contributing equally with  a weight
$2^{1-N_K}$.

Comparing  the remaining  test sources  and functional  derivatives in
eq.~(\ref{eq:gengamma3}),  the number  of  $+$ type  outgoing legs  is
given by
\begin{equation}
  N_K^{+}\ =\ P\:-\:2V^+\ =\ 2(N\:-\:V^+)\:-\:3 \ \leq\ N_K\;.
\end{equation}
Notice that, for $\phi^3$ theory,  $N_K^+$ is always odd. Thus the set
of graphs consistent with a given topology corresponds to the sum over
all  ways of  drawing that  topology with  an odd  number of  $+$ type
outgoing legs. The  number of ways of arranging  $N_K^+$ outgoing legs
in  a graph  with  a total  of  $N_K$ outgoing  legs  is the  binomial
coefficient:
\begin{equation}
  \frac{N_K!}{N_K^{+}!(N_K\:-\:N_K^+)!}\ =\ \frac{N_K!}{N_K^+!N_K^-!}\;.
\end{equation}
The total  number of graphs consistent  with a given  topology is then
obtained by summing over all odd $1\:\leq\:N_K^+\:\leq\:N_K$, i.e.
\begin{equation}
  \sum_{\substack{N_K^+\:\geq\:1 \\ \mathrm{odd}}}^{N_K}
  \frac{N_K!}{N_K^+!(N_K\:-\:N_K^+)!}\ =\ 2^{N_K-1}\;.
\end{equation}
This  factor exactly  cancels  the overall  factor  of $2^{1-N_K}$  in
$\Gamma_{JK}$ and we  are left with a sum  over tree-level topologies,
each with unit weight and entirely equivalent to the S-matrix result.

This   equivalence  with   the   S-matrix  can   be  extended   beyond
tree-level.   Specifically,  a   general  retarded   Green's  function
($\Gamma^{n \to m}_{\mathrm{R}}$) can  be obtained by summing over all
circlings  (see  appendix A)  except  those  of  the $m$  largest-time
points, i.e.
\begin{equation}
  \Gamma^{n \to m}_{\mathrm{R}} = \Gamma^{n \to m}_{\mathrm{F}} +
  \sum_{\text{circlings} \; \odot} \Gamma^{n \to m}_{\odot}
  \label{eq:largest-time}
\end{equation}
and we have isolated  the zero-circlings contribution corresponding to
a graph built entirely  from Feynman propagators. Since the scattering
amplitude    $\Gamma^{n    \to    m}_{JK}$,   derived    in    section
\ref{sec:pathintegral},  can be  obtained from  this  retarded Green's
function  after  convoluting with  the  source/detector functions,  it
follows that the same $\Gamma^{n  \to m}_{JK}$ could be obtained using
the   corresponding    Feynman   Green   function,    $\Gamma^{n   \to
m}_{\mathrm{F}}$. This  is because the  second term on  the right-hand
side  of eq.~(\ref{eq:largest-time})  vanishes if  we impose  that the
incoming particles  carry positive energy  forwards in time,  which is
the case  when evaluating  the S-matrix. At  tree-level, we  have just
shown that the combinatoric  factor associated with the convolution of
the  Green's function  and  the  sources is  exactly  as required  for
agreement with the  S-matrix result, and this is  sufficient to insure
equivalence at all orders.

\paragraph{Note  added:} Whilst  preparing the  final version  of this
manuscript  we became  aware of  reference  \cite{Musso:2006pt}, which
presents  a  diagrammatic approach  to  the  calculation of  ``in-in''
expectation   values   similar    to   that   presented   in   section
\ref{sec:expectation} in the absence of external sources.

\acknowledgments

We should like  to thank Ed Copeland, Fay  Dowker, Tim Hollowood, Leif
L\"onnblad,  Tim  Morris,  Mike  Seymour  and Graham  Shore  for  many
enjoyable  and helpful  discussions. We  also thank  Sean  Carroll for
provoking us in the first  place.  This work is partially supported by
the Lancaster-Manchester-Sheffield  Consortium for Fundamental Physics
under STFC grant ST/J000418/1 and by the Royal Society. The work of PM
is supported in  part by the IPPP through  STFC grant ST/G000905/1. PM
would like to acknowledge the conferment of visiting researcher status
at the University of Sheffield.

\appendix

\section{Relation to unitarity cutting rules}
\label{sec:cutting}

We will  now illustrate that the manifestly  causal amplitudes derived
from  the  operator and  path-integral  approaches  are precisely  the
retarded amplitudes obtained by  means of the Kobes-Semenoff unitarity
cutting   rules   of   the  ``in-in"   formalism   \cite{Kobes:1985kc,
Kobes:1986za}. For the purposes of this section, we will omit to write
the  physical sources  and convolution  integrals associated  with the
external propagators and we  take the limits $T_A\:\to\:-\:\infty$, so
that interaction vertices are integrated over an infinite domain.

We begin  by noting the  following diagrammatic representation  of the
time-ordered   (Feynman),  anti-time-ordered   (Dyson)   and  Wightman
propagators (e.g.~see \cite{Veltman:1994wz}):
\begin{subequations}
  \label{eq::propsdiag}
  \begin{align}
    \Delta_{\mathrm{F}}(x,y)\ &=\ \ \parbox[c]{6em}{\includegraphics[]{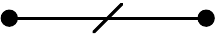}}\\
    \Delta_{\mathrm{D}}(x,y)\ &=\ \parbox[c]{6em}{\includegraphics[]{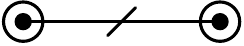}}
    \hspace{0.75em} =\ \Delta_{\mathrm{F}}^*(x,y)\\
    -\:\Delta_{>}(x,y)\ &=\ \parbox[c]{6em}{\includegraphics[]{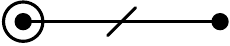}}
    \hspace{0.75em}
    =\ -\:\Delta_<^* (x,y)\\
    -\:\Delta_{<}(x,y)\ &=\
    \ \parbox[c]{6em}{\includegraphics[]{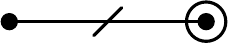}}\hspace{0.75em}.
  \end{align}
\end{subequations}
We  can follow the  energy flow  through a  general graph  built using
these propagators since positive energy always flows from an uncircled
into a circled vertex. The propagators satisfy
\begin{equation}
  \parbox[c]{5.5em}{\includegraphics[]{F.pdf}}\hspace{0.75em}
  +\ \parbox[c]{6em}{\includegraphics[]{D.pdf}}\hspace{0.75em}
  +\ \parbox[c]{5.5em}{\includegraphics[]{G.pdf}}\hspace{0.75em}
  +\ \parbox[c]{6em}{\includegraphics[]{L.pdf}}\ =\ 0\;.
\end{equation}
Notice  that, unlike  the  usual unitarity  cutting  rules applied  to
S-matrix  theory, the  Kobes-Semenoff  cutting rules  do not  restrict
diagrams to contain only positive  energy flow. As we shall see below,
negative  energy flow is  necessary for  the construction  of retarded
diagrams and the restoration of manifest causality.

In  terms of  the propagators  above, the  retarded propagator  can be
expressed as
\begin{align}
  \label{eq::retdiag}
  \:\Delta_{\mathrm{R}}(x,y)\
  \equiv \parbox[c]{5.5em}{\includegraphics[]{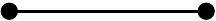}}\ &= \
  \ \parbox[c]{5.5em}{\includegraphics[]{F.pdf}}\hspace{0.75em}+\
  \ \parbox[c]{6em}{\includegraphics[]{L.pdf}}\nonumber\\
  &= \ -\ \parbox[c]{6em}{\includegraphics[]{D.pdf}}\hspace{0.75em}
  -\ \parbox[c]{5.5em}{\includegraphics[]{G.pdf}}\nonumber\\
  &= \hspace{0.5em} \parbox[c]{6em}{\includegraphics[]{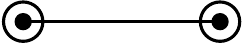}}\;
  \nonumber \\
  &= \ - \ \parbox[c]{6em}{\includegraphics[]{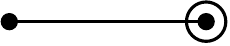}} \ =\ 
  - \ \parbox[c]{6em}{\includegraphics[]{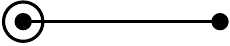}}\;.
\end{align}
In other words, circling plays no role for retarded propagators except
to help keep track of the minus signs.

By  virtue   of  the  Kobes-Semenoff  cutting   rules,  the  one-loop,
negative-frequency  Wightman propagator  is obtained  by  circling the
right-most external vertex and  summing over all possible circlings of
the internal vertices. The one-loop retarded propagator then takes the
form
\begin{align}
  \label{eq::oneloopR}
  \Delta_{\mathrm{R}}^{(1)}(x,y)\ &=\ \Delta_{\mathrm{F}}^{(1)}(x,y)\:
  -\:\Delta_{<}^{(1)}(x,y)\nonumber\\&
  =\hspace{0.5em} \parbox[c]{5.5em}{\includegraphics[]{F.pdf}}\hspace{0.75em}
  +\ \parbox[c]{6em}{\includegraphics[]{L.pdf}}\nonumber\\&
  +\ \parbox[c]{10.5em}{\includegraphics[]{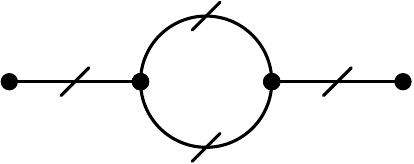}}\nonumber\\&
  +\ \parbox[c]{10.5em}{\includegraphics[]{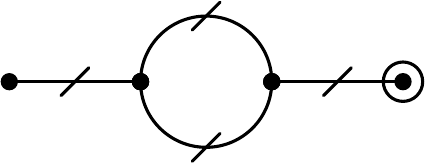}}\hspace{1em}
  +\ \parbox[c]{10.5em}{\includegraphics[]{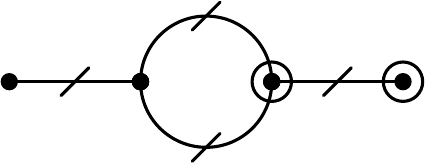}}\nonumber\\&
  +\ \parbox[c]{10.5em}{\includegraphics[]{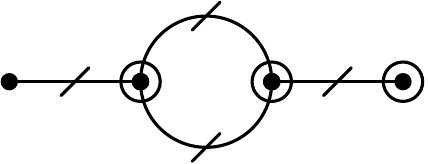}}\hspace{1em}
  +\ \parbox[c]{10.5em}{\includegraphics[]{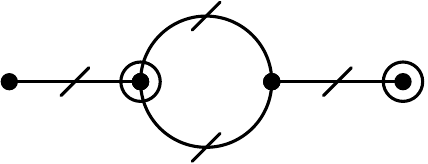}}\hspace{1.25em}.
\end{align}
Since energy  is conserved  through the internal  vertices the following
circlings are identically zero:
\begin{equation}
  \label{eq::oneloopzeros}
  \parbox[c]{10.5em}{\includegraphics[]{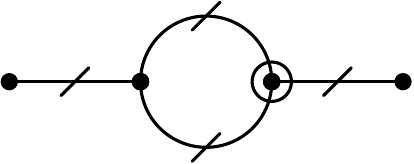}}\hspace{0.5em}
  =\hspace{0.25em} \parbox[c]{10.5em}{\includegraphics[]{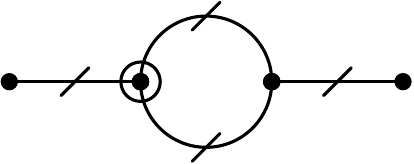}}
  \hspace{0.5em}
  =\hspace{0.25em} \parbox[c]{10.5em}{\includegraphics[]{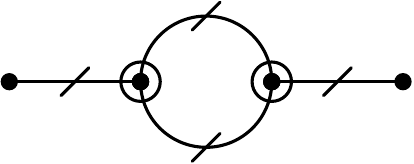}}
  \hspace{0.5em} =\ 0\;.
\end{equation}
Notice  that  the last  diagram  in  eq.~(\ref{eq::oneloopR}) is  also
vanishing.  Adding to  eq.~(\ref{eq::oneloopR}) the vanishing diagrams
from eq.~(\ref{eq::oneloopzeros}), we  make the following observation:
the  retarded  diagram  is  obtained  by  summing  over  all  possible
circlings  of vertices  whilst  leaving the  outgoing, leftmost  point
uncircled \cite{Kobes:1990ua}.

The one-loop contribution then contains 8 diagrams:
\begin{align}
  \label{eq::oneloop8}
  \Delta_{\mathrm{R}}^{(\delta 1)}(x,y)\ &=\
  \parbox[c]{10.5em}{\includegraphics[]{oneselfF.pdf}}\hspace{0.75em}
  +\ \parbox[c]{10.5em}{\includegraphics[]{oneselfL1.pdf}}\nonumber\\ &
  +\ \parbox[c]{10.5em}{\includegraphics[]{oneselfL2.pdf}}\hspace{1em}
  +\ \parbox[c]{10.5em}{\includegraphics[]{oneselfZ1.pdf}}\nonumber\\&
  +\ \parbox[c]{10.5em}{\includegraphics[]{oneselfL3.pdf}}\hspace{1em}
  +\ \parbox[c]{10.5em}{\includegraphics[]{oneselfZ3.pdf}}\nonumber\\&
  +\ \parbox[c]{10.5em}{\includegraphics[]{oneselfL4.pdf}}\hspace{1em}
  +\ \parbox[c]{10.5em}{\includegraphics[]{oneselfZ2.pdf}}\hspace{0.75em},
\end{align}
where                                     $\Delta_{\mathrm{R}}^{(\delta
1)}(x,y)\:=\:\Delta_{\mathrm{R}}^{(1)}(x,y)\:-\:\Delta_{\mathrm{R}}(x,y)$.
Combining    each    pair    of     diagrams    row    by    row    in
eq.~(\ref{eq::oneloop8}),      using       the      identities      in
eq.~(\ref{eq::retdiag}), we obtain
\begin{align}
  \Delta_{\mathrm{R}}^{(\delta 1)}(x,y)\ &=\
  \parbox[c]{10.5em}{\includegraphics[]{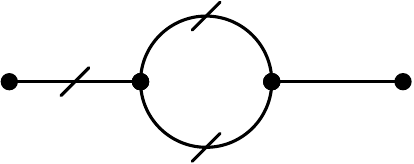}}\hspace{0.75em}
  +\ \parbox[c]{10.5em}{\includegraphics[]{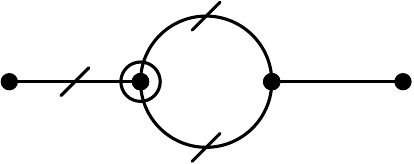}}\nonumber\\&
  +\ \parbox[c]{10.5em}{\includegraphics[]{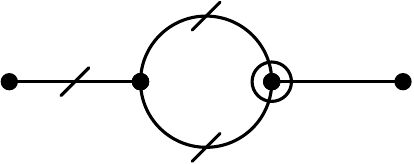}}\hspace{0.75em}
  +\ \parbox[c]{10.5em}{\includegraphics[]{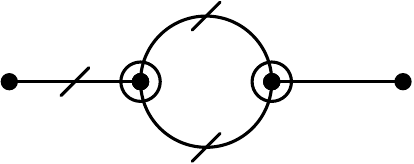}}\hspace{0.75em}.
\end{align}
Splitting the remaining diagrams into their component parts, we may
write
\begin{align}
  \Delta_{\mathrm{R}}^{(\delta 1)}(x,y)\ &=\
  \parbox[c]{5.85em}{\includegraphics[]{F.pdf}}
  \bigg[\parbox[c]{5.5em}{\includegraphics[]{F.pdf}\\
    \includegraphics[]{F.pdf}}\:
  -\:\parbox[c]{6em}{\includegraphics[]{L.pdf}\\
    \includegraphics[]{L.pdf}}\bigg]\
  \parbox[c]{5.5em}{\includegraphics[]{RF.pdf}}\nonumber\\&
  +\ \parbox[c]{6em}{\includegraphics[]{L.pdf}}
  \bigg[\parbox[c]{6em}{\includegraphics[]{G.pdf}\\
    \parbox[c]{0.0em}{\ }\includegraphics[]{G.pdf}}\:
  -\:\parbox[c]{6em}{\includegraphics[]{D.pdf}\\
    \includegraphics[]{D.pdf}}\ \bigg]\
  \parbox[c]{6em}{\includegraphics[]{RF.pdf}}\;.
\end{align}
Again using the identities in eq.~(\ref{eq::retdiag}), this may recast
in terms of Feynman and retarded propagators as
\begin{align}
  \Delta_{\mathrm{R}}^{(\delta 1)}(x,y)\ &=\
  \parbox[c]{5.5em}{\includegraphics[]{RF.pdf}}\
  \bigg[2\ \parbox[c]{5.5em}{\includegraphics[]{F.pdf}\\
    \includegraphics[]{RF.pdf}}\:
  -\:\parbox[c]{5.5em}{\includegraphics[]{RF.pdf}\\
    \includegraphics[]{RF.pdf}}\ \bigg]\
  \parbox[c]{6em}{\includegraphics[]{RF.pdf}}\;.
\end{align}
Finally,  after recombining  the component  pieces, we  arrive  at the
one-loop retarded propagator
\begin{equation}
  \Delta_{\mathrm{R}}^{(1)}(x,y)\ =\
  \parbox[c]{5.5em}{\includegraphics[]{RF.pdf}}\hspace{0.75em}
  +\ 2\ \parbox[c]{10.5em}{\includegraphics[]{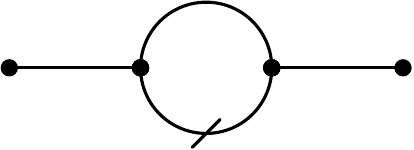}}\hspace{0.75em}
  -\ \parbox[c]{10.5em}{\includegraphics[]{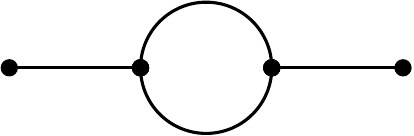}}
\end{equation}
which is of precisely the expected form, i.e.
\begin{equation}
  \Delta_{\mathrm{R}}^{(1)}(x,y)\ =\ \Delta_{\mathrm{R}}(x,y)\:
  +\:\Delta_{\mathrm{R}}(x,z)\star\Pi_{\mathrm{R}}^{(1)}(z,z')
  \star\Delta_{\mathrm{R}}(z',y)\;,
\end{equation}
where 
\beq
  \Pi_{\mathrm{R}}^{(1)}(z,z') =
  \frac{(-ig)^2}{2!}\Big[2\Delta_{\mathrm{F}}(z,z')
  \Delta_{\mathrm{R}}(z,z')\:
  -\:\big(\Delta_{\mathrm{R}}(z,z')\big)^2\Big]\;
\eeq 
is  the truncated one-loop  retarded self  energy and  $\star$ denotes
integration over the intermediate spacetime points $z$ and $z'$.

It is  interesting to see how  this works starting  from the truncated
self-energy.  In terms of the Kobes-Semenoff cutting rules, this is
\begin{align}
  \Pi_{\mathrm{R}}^{(1)}(z,z')\ & =\ \Pi^{(1)}(z,z')\:
  -\:\Pi^{(1)}_{<}(z,z')\ = \frac{(-ig)^2}{2!}
  \Big[\big(\Delta_{\mathrm{F}}(z,z')\big)^2\:
  -\:\big(\Delta_<(z,z')\big)^2\Big]
  \nonumber\\ &
  =\ \parbox[c]{3.75em}{\includegraphics[]{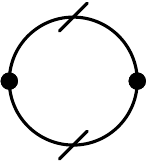}}\:
  +\:\parbox[c]{3.75em}{\includegraphics[]{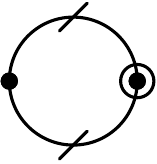}}\ \ .
\end{align}
Proceeding as before by  separating the truncated self-energy into its
component parts, we obtain
\begin{align}
  \Pi_{\mathrm{R}}^{(1)}(z,z')\ &=\
  \parbox[c]{0.75em}{\includegraphics[]{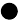}}
  \bigg[\parbox[c]{5.5em}{\includegraphics[]{F.pdf}\\
    \includegraphics[]{F.pdf}}\:
  -\:\parbox[c]{6em}{\includegraphics[]{L.pdf}\\
    \includegraphics[]{L.pdf}}\bigg]\
  \parbox[c]{1em}{\includegraphics[]{uncircle.pdf}}\;.
\end{align}
Substituting  the  decomposition   of  the  retarded  propagator  from
eq.~(\ref{eq::retdiag}), this may be  written in terms of only Feynman
and retarded propagators exactly as before, i.e.
\begin{align}
  \Pi_{\mathrm{R}}^{(1)}(z,z')\ &=\
  2\ \parbox[c]{3.75em}{\includegraphics[]{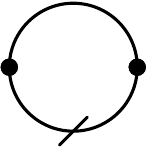}}\:
  -\:\parbox[c]{3.75em}{\includegraphics[]{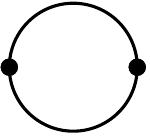}}~.
\end{align}

Following  the  same circling  rules  for  the one-to-two  scattering,
i.e.~we do not circle  the ``latest time'' vertices that we anticipate
coupling  to detector sources  $K$, the  retarded contribution  to the
time-ordered 3-point function is given by
\begin{align}
  \Gamma^{1 \to 2}_{\mathrm{R}}(x,y,z)\
  &=\ \parbox[c]{3.75em}{\includegraphics[]{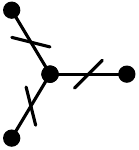}}\:
  +\:\parbox[c]{4em}{\includegraphics[]{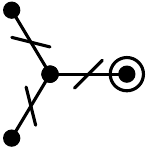}}\:
  +\: \parbox[c]{3.75em}{\includegraphics[]{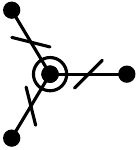}}\:
  +\:\parbox[c]{4em}{\includegraphics[]{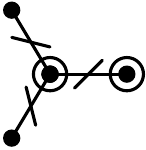}}
  \nonumber\\&
  = \ \parbox[c]{3.75em}{\includegraphics[]{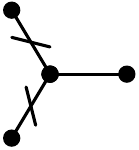}}\:
  +\:\parbox[c]{4em}{\includegraphics[]{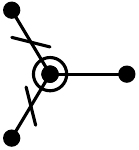}}\;.
\end{align}
Separating the component contributions, we have
\begin{align}
  \Gamma^{1\to 2}_{\mathrm{R}}(x,y,z)\ &=\
  \bigg[\parbox[c]{5.5em}{\includegraphics[]{F.pdf}\\
    \includegraphics[]{F.pdf}}\:
  -\:\parbox[c]{6em}{\includegraphics[]{L.pdf}\\
    \includegraphics[]{L.pdf}}\bigg]\
  \parbox[c]{6em}{\includegraphics[]{RF.pdf}}\;,
\end{align}
yielding
\begin{align}
  \Gamma^{1 \to 2}_{\mathrm{R}}(x,y,z)\
  &=\ \parbox[c]{4em}{\includegraphics[angle=0]{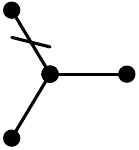}}\: +
  \: \parbox[c]{4em}{\includegraphics[]{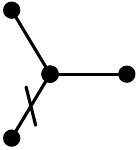}}\:
  -\:\parbox[c]{4em}{\includegraphics[]{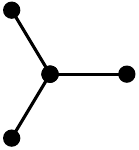}}\;,
\end{align}
which is in agreement with eq.~(\ref{eq:1to2p}) after convoluting with
the source functions.

We   may  repeat   this  diagrammatic   manipulation   for  two-to-two
scattering, which corresponds to 16 circlings:
\begin{align}
  \Gamma^{2\to 2}_{\mathrm{R}}(x,y,z,w)\ &=\
  \parbox[c]{6em}{\includegraphics[]{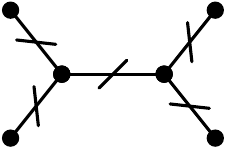}}\:
  +\:\parbox[c]{6em}{\includegraphics[]{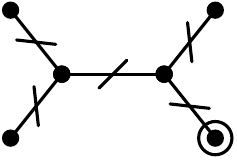}}\:
  +\:\parbox[c]{6em}{\includegraphics[]{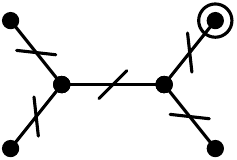}}\:
  +\:\parbox[c]{6em}{\includegraphics[]{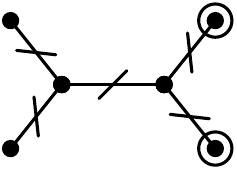}}
  \nonumber\\&
  +\:\parbox[c]{6em}{\includegraphics[]{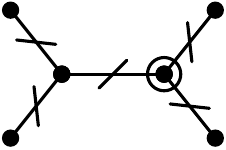}}\:
  +\:\parbox[c]{6em}{\includegraphics[]{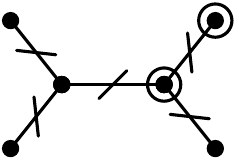}}\:
  +\:\parbox[c]{6em}{\includegraphics[]{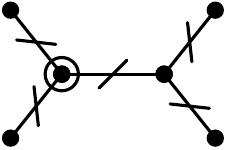}}\:
  +\:\parbox[c]{6em}{\includegraphics[]{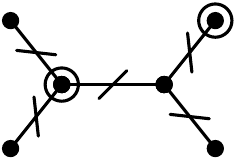}}
  \nonumber\\&+\:\parbox[c]{6em}{\includegraphics[]{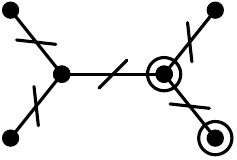}}\:
  +\:\parbox[c]{6em}{\includegraphics[]{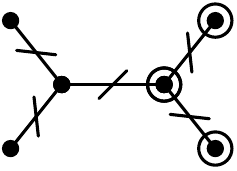}}\:
  +\:\parbox[c]{6em}{\includegraphics[]{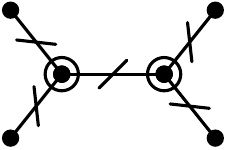}}\:
  +\:\parbox[c]{6em}{\includegraphics[]{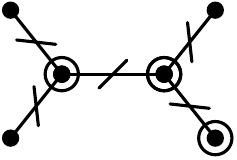}}
  \nonumber\\&
  +\parbox[c]{6em}{\includegraphics[]{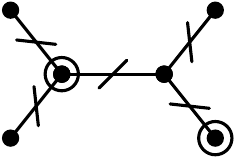}}\:
  +\:\parbox[c]{6em}{\includegraphics[]{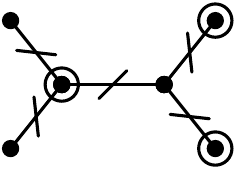}}\:
  +\:\parbox[c]{6em}{\includegraphics[]{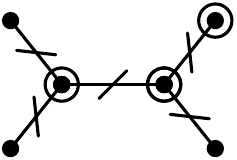}}\:
  +\:\parbox[c]{6em}{\includegraphics[]{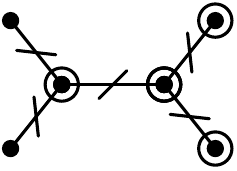}}\;.
\end{align}
After pairwise  contracting the  diagrams and expanding  the component
propagators  by means  of  eq.~(\ref{eq::retdiag}), we  may show  that
these 16 ordered diagrams reduce to the following three diagrams:
\begin{equation}
  \Gamma^{2 \to 2}_{\mathrm{R}}(x,y,z,w)\
  =\parbox[c]{6em}{\includegraphics[]{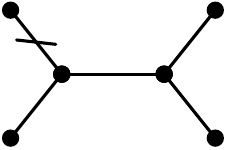}}\:
  +\: \parbox[c]{6em}{\includegraphics[]{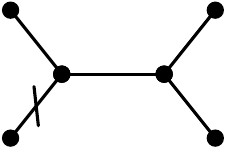}}\:
  -\:\parbox[c]{6em}{\includegraphics[]{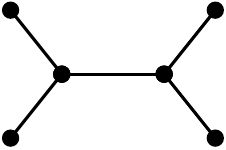}}\;,
\end{equation}
again in  agreement with the  earlier results (see the  sentence below
eq.~(\ref{eq:smat})). The results of  this section serve to illustrate
the role played  by negative energy flow forwards  in time in building
causal amplitudes.

\section{Propagator definitions}
\label{app:props}

\noindent  Here we  collect together  the definitions  of  the various
propagators that appear.

\begin{itemize}

\item The Wightman propagators:
  \begin{align}
    \Delta_>(x,y) &\ =\  \langle \phi(x) \phi(y) \rangle =
    \int \frac{\dd^3 \bs{p}}{(2\pi)^3 2E} e^{-ip \cdot (x-y)}
    \nonumber\\ &\ =\ \int \frac{\dd^4 p}{(2\pi)^3}
    \delta(p^2 - m^2) \Theta(p_0) e^{-ip \cdot (x-y)}~,
   \displaybreak \\ 
   \Delta_<(x,y) &\ =\  \langle \phi(y) \phi(x) \rangle =
   \int \frac{\dd^3 \bs{p}}{(2\pi)^3 2E} e^{+ip \cdot (x-y)}
   \nonumber\\ &\ =\ \int \frac{\dd^4 p}{(2\pi)^3}
   \delta(p^2 - m^2) \Theta(-p_0) e^{-ip \cdot (x-y)}~.
\end{align}

\item The Pauli-Jordan propagator,
  $\Delta(x,y) = \Delta_>(x,y) - \Delta_<(x,y)$:
  \beq
  \Delta(x,y) = \Delta_{yx} = \langle [\phi(x),\phi(y)] \rangle
  = \int \frac{\dd^3 \bs{p}}{(2\pi)^3 2E} \left( e^{-ip \cdot (x-y)}
    -  e^{+ip \cdot (x-y)}\right)~.
\eeq

\item The Hadamard propagator,
  $\Delta_1(x,y) = \Delta_>(x,y) + \Delta_<(x,y)$:
  \beq
  \Delta_1(x,y) = \langle \{\phi(x),\phi(y)\} \rangle
  = \int \frac{\dd^3 \bs{p}}{(2\pi)^3 2E} \left( e^{-ip \cdot (x-y)}
    +  e^{+ip \cdot (x-y)} \right)~.
\eeq

\item The Feynman and Dyson propagators,
  $\Delta_\text{F}(x,y) = \Delta^*_\text{D}(x,y)$:
  \bea
    \Delta_\text{F}(x,y) &=& \langle T[\phi(x) \phi(y)]\rangle
    = \Delta_>(x,y) \Theta(x^0-y^0) + \Delta_<(x,y) \Theta(y^0-x^0)
    \nonumber \\
    &=& \int \frac{\dd^4 k}{(2\pi)^4} e^{-ik\cdot(x-y)}
    \frac{i}{k^2-m^2+i \epsilon}
    \nonumber \\
    &=& i \int \frac{\dd^4 k}{(2\pi)^4} e^{-ik\cdot(x-y)}
    \left[P\left(\frac{1}{k^2-m^2}\right)-i \pi \delta(k^2-m^2) \right]
    \eea
    where $P$ denotes the Cauchy Principal Value.

\item The retarded and advanced propagators,
  $\Delta_\text{R}(x,y) = \Delta_\text{A}(y,x)$:
  \bea
    \Delta_\text{R}(x,y) &=&  \Delta(x,y) \Theta(x^0-y^0)
    \nonumber \\
    &=& \int \frac{\dd^4 k}{(2\pi)^4} e^{-ik\cdot(x-y)}
    \frac{i}{(k_0+i\epsilon)^2-\bs{k}^2-m^2}
    \nonumber \\ &=& i \int \frac{\dd^4 k}{(2\pi)^4} e^{-ik\cdot(x-y)}
    \left[P\left(\frac{1}{k^2-m^2}\right)-i \pi \delta(k^2-m^2)
      \text{sgn}(k_0)\right]~.
    \eea
  \end{itemize}

\bibliography{causality}
\bibliographystyle{JHEP}

\end{document}